\documentclass[twocolumn,floats,amsmath,amssymb,nofootinbib]{aastex63}
\usepackage{bm}
\usepackage{hyperref}
\usepackage[english]{babel}
\usepackage{tabu}
\usepackage{amsmath}
\usepackage[caption=false, position=t, singlelinecheck=off]{subfig}
\usepackage[T1]{fontenc} 
\usepackage{multirow}
\usepackage[encapsulated]{CJK}
\usepackage{ucs}
\usepackage[utf8x]{inputenc}

\newcommand{\cntext}[1]{\begin{CJK}{UTF8}{bkai}#1\ignorespacesafterend\end{CJK}} 

\received{}
\revised{}
\accepted{}
\submitjournal{ApJ}

\shorttitle{Embryo Formation in Self-Gravitating Disks}
\shortauthors{Baehr et al. 2022}

\graphicspath{{./}{figures/}}

\makeatletter
\newcommand*\bigcdot{\mathpalette\bigcdot@{.5}}
\newcommand*\bigcdot@[2]{\mathbin{\vcenter{\hbox{\scalebox{#2}{$\m@th#1\bullet$}}}}}
\makeatother

\begin{document}

\title{Direct Formation of Planetary Embryos in Self-Gravitating Disks}

\citestyle{egu}
\bibliographystyle{yahapj}

\correspondingauthor{Hans Baehr}
\email{hans-paul.baehr@unlv.edu}

\author[0000-0002-0880-8296]{Hans Baehr}
\affil{Department of Physics and Astronomy, University of Nevada, Las Vegas, 4505 South Maryland Parkway, Las Vegas, NV 89154, USA}
\affil{Nevada Center for Astrophysics, University of Nevada, Las Vegas, 4505 South Maryland Parkway, Las Vegas, NV 89154, USA}

\author[0000-0003-3616-6822]{Zhaohuan Zhu (\cntext{朱照寰})}
\affil{Department of Physics and Astronomy, University of Nevada, Las Vegas, 4505 South Maryland Parkway, Las Vegas, NV 89154, USA}
\affil{Nevada Center for Astrophysics, University of Nevada, Las Vegas, 4505 South Maryland Parkway, Las Vegas, NV 89154, USA}

\author[0000-0003-2589-5034]{Chao-Chin Yang (\cntext{楊朝欽})}
\affil{Department of Physics and Astronomy, University of Nevada, Las Vegas, 4505 South Maryland Parkway, Las Vegas, NV 89154, USA}
\affil{Nevada Center for Astrophysics, University of Nevada, Las Vegas, 4505 South Maryland Parkway, Las Vegas, NV 89154, USA}

\begin{abstract}
Giant planets have been discovered at large separations from the central star. Moreover, a striking number of young circumstellar disks have gas and/or dust gaps at large orbital separations, potentially driven by embedded planetary objects. To form massive planets at large orbital separations through core accretion within disk lifetime, however, an early solid body to seed pebble and gas accretion is desirable. Young protoplanetary disks are likely self-gravitating, and these gravitoturbulent disks may efficiently concentrate solid material at the midplane driven by spiral waves. We run 3D local hydrodynamical simulations of gravitoturbulent disks with Lagrangian dust particles to determine whether particle and gas self-gravity can lead to the formation of dense solid bodies, seeding later planet formation. When self-gravity between dust particles is included, solids of size $\mathrm{St} = 0.1$ to $1$ concentrate within the gravitoturbulent spiral features and collapse under their own self-gravity into dense clumps up to several $M_{\oplus}$ in mass at wide orbits. Simulations with dust that drift most efficiently, $\mathrm{St}=1$, form the most massive clouds of particles, while simulations with smaller dust particles, $\mathrm{St}=0.1$, have clumps with masses an order of magnitude lower.  When the effect of dust backreaction onto the gas is included, dust clumps become smaller by a factor of a few but more numerous. The existence of large solid bodies at an early stage of the disk can accelerate the planet formation process, particularly at wide orbital separations, and potentially explain planets distant from the central stars and young protoplanetary disks with substructures.
\end{abstract}


\section{Introduction}
\label{sec:intro}

Observing young stars and disks still embedded in their natal cores is challenging since the progenitor envelope is opaque at most wavelengths. This makes it difficult to constrain the initial conditions of planet formation, including the early sedimentation and concentration of dust. During this early time, the disk is potentially massive enough to be gravitationally unstable. Depending on the cooling efficiency, gravitationally unstable disks can either fragment into large, bound, gas-dominated companions \citep{Boss1997,Gammie2001,Meru2011a} or become marginally stable to form large scale spiral structures \citep{Mejia2005,Zhu2012,Kratter2016}. While observing the dynamics and processes of gravitationally unstable disks is difficult due to their rapid accretion and thus short lifespans, they can play an important role on subsequent planet formation \citep{Forgan2019}.

\begin{deluxetable*}{cccccccccccccc}
\tablecaption{List of Simulations:\label{tab:sims}}
\tablehead{
\colhead{Model} & \colhead{$\beta$} & \colhead{ \begin{tabular}[c]{@{}l@{}} Particle \\  backreaction? \end{tabular}} & \colhead{$\mathrm{St}$} & \colhead{$Q$} & \colhead{$\alpha_R$} & \colhead{$\alpha_G$} & \colhead{$H_{d} [H_{g}]$} & \colhead{$\delta_{d,x}$}& \colhead{$\delta_{d,z}$} & \colhead{$\mathrm{Sc}$} & \colhead{ \begin{tabular}[c]{@{}l@{}} $\sigma_{d,x}$ \\  $[c_s]$ \end{tabular}} & \colhead{\begin{tabular}[c]{@{}l@{}} $\sigma_{d,z}$ \\  $[c_s]$ \end{tabular}}}
\startdata
noBR\_S\_10 & $10$ & No & $0.1$ & $1.2$ & $4.9\times 10^{-3}$ & $0.03$ & $9.8\times 10^{-2}$ & $4.5\times 10^{-3}$ & $9.6\times 10^{-4}$ & $36$  & $0.36$ & $0.09$\\
noBR\_L\_10  & $10$ & No & $1$ & $1.3$ & $6.5\times 10^{-3}$ & $0.044$ & $6.7\times 10^{-3}$ & $2.5\times 10^{-2}$ & $4.5\times 10^{-5}$ & $113$  & $0.56$ & $0.08$\\
BR\_S\_10 & $10$ & Yes & $0.1$ & $1.3$ & $7.9\times 10^{-3}$ & $0.029$ & $9.1\times 10^{-2}$ & $5.6\times 10^{-3}$ & $8.4\times 10^{-4}$ & $44$ & $0.43$ & $0.11$ \\
BR\_L\_10 & $10$ & Yes & $1$ & $1.3$ & $6.3\times 10^{-3}$ & $0.039$ & $9.3\times 10^{-3}$ & $2.1\times 10^{-2}$ & $8.6\times 10^{-5}$ & $406$  & $0.72$ & $0.08$\\
BR\_XL\_10& $10$ & Yes & $10$ & $1.3$ & $6.0\times 10^{-3}$ & $0.021$ & $5.8\times 10^{-2}$ & $1.1\times 10^{-2}$ & $1.2\times 10^{-4}$ & $225$ & $0.87$ & $0.07$ \\
BR\_L\_5 & $5$ & Yes & $1$ & $1.4$ & $9.7\times 10^{-3}$ & $0.11$ & $1.1\times 10^{-2}$  & $1.9\times 10^{-2}$ & $3.4\times 10^{-2}$ & $3.5$ & $0.94$ & $0.09$ \\
noBR\_L\_10\_HR & $10$ & No & $1$ & $1.4$ & $6.8\times 10^{-3}$ & $0.048$ & $1.1\times 10^{-2}$  & $1.4\times 10^{-2}$ & $1.2\times 10^{-4}$ & $458$ & $0.66$ & $0.09$ \\
\enddata
\tablecomments{Simulation parameters and steady-state values of the stability parameter $Q$, Reynolds' stress $\alpha_R$, gravitational stress $\alpha_G$, dust scale height $H_{d}$, dimensionless particle diffusion constants $\delta$ and particle velocity dispersions $\sigma$. We define the Schmidt number $\mathrm{Sc} \equiv (\alpha_R + \alpha_G)/\delta_{d,z}$. All simulations have spatial resolution $512^{2}\times 256$ with box lengths $L_{x}=L_{y}=(80/\pi) H_g$ and $L_{z}=(40/\pi) H_g$, except for the high resolution simulation which used $1024^{2}\times 512$ cells}. Simulations are initially marginally gravitationally stable such that the gas $Q_{0} = 1.02$ and the solid-to-gas mass ratio $Z_0 = M_{d}/M_{g} = 0.01$. For measured quantities within the particle clouds, see Table \ref{tab:clumps}. 
\end{deluxetable*}

Recent ALMA protoplanetary disk observations revealed that substructure forms early and often, such that rings are nearly ubiquitous and spirals are occasionally present as well \citep{Andrews2018a,Long2018,Clarke2018,Segura-Cox2020,Sheehan2020}. Although the mechanisms to form rings and gaps are heavily debated, the presence of undetected planets which can carve open these gaps is a notable option \citep{Zhang2018,Lodato2019,Wang2021a}, a scenario being tested by gas kinematic observations \citep{Pinte2018,Teague2021}. On the other hand, the existence of systems with multiple giant planets on wide orbits, such as HR 8799 \citep{Marois2010,Maire2016} suggests that massive planets can indeed form far away from the central star. Thus, it is natural to ask how to form planets at large orbital separations where core and pebble accretion models are less efficient. Planetesimals and even planetary cores may have formed before the Class II stage and could have begun accreting gaseous envelopes and opening gaps in disks.

The fragmentation of gravitationally unstable disks is sometimes invoked as a mechanism to directly form giant planets at distant orbital separations \citep{Cheetham2018,Bonnefoy2018,Morales2019,Janson2021}, with initial masses expected to be at least a few Jupiter masses \citep{Boss1997,Baehr2017}. However, the planets that potentially inhabit the gaps of young disks are typically sub-Jovian, likely too low to be explained by direct fragmentation of the gas disk \citep{Zhang2018,Lodato2019}. Tidal downsizing of companions formed by disk fragmentation could explain the smaller planets at closer radii, but, at orbital separations greater than $\sim$10 au, the tidal force cannot strip enough material \citep{Nayakshin2010}. Thus, the planets may still grow in the traditional ``core-accretion'' fashion, starting with dense solid embryos or cores that only accrete substantial gas envelopes later. Self-gravitating disks which do not fragment but are instead marginally unstable (a.k.a. gravitoturbulent) could potentially concentrate solid material enough so that the dust clouds are gravitationally bound and form planetary cores/embryos directly \citep{Rice2004,Boley2010}.

Marginally gravitationally unstable disks occur when the gravitational stability parameter \citep{Safronov1960,Toomre1964,Goldreich1965}
\begin{equation}
Q = \frac{c_{\mathrm{s}}\Omega}{\pi G \Sigma},
\end{equation}
is slightly above unity. With $G$ being the gravitational constant, stability from thermal pressure and rotational shear is quantified through the gas sound speed $c_{\mathrm{s}}$ and Keplerian orbital frequency $\Omega$, respectively. These counteract gravitational collapse of perturbations in a disk with gas surface density $\Sigma$. In addition, if the cooling timescale $t_\mathrm{c}$ is long enough, $t_{\mathrm{c}} > \beta\Omega^{-1}$ \citep{Gammie2001}, fragmentation of the disk into dense gas structures may be prevented. There remains some uncertainty about the value of $\beta$ and whether this is a sufficient criterion \citep{Paardekooper2011,Meru2012,Brucy2021}. In 3D simulations like those in this work, a criterion of $\beta = 3$ has been shown to converge with resolution \citep{Baehr2017}. Thus, we focus on gravitoturbulent disks with $\beta > 3$.

Spiral arms are generated in these gravitoturbulent disks \citep{Cossins2009}. These spirals
can concentrate solids and potentially produce planetesimals, as shown in both local \citep{Gibbons2012,Gibbons2014a,Shi2016} and global simulations \citep{Rice2004,Boley2010,Booth2016,Cadman2020a}. The studies using local simulations were all two-dimensional which did not consider how dust settling affects the formation of a thin dust layer. The global studies are all in 3D, but did not include self-gravity of the dust, which is necessary for the formation of self-bound dust clouds. Even so, these studies showed that gravitoturbulent disks could potentially provide young disks with the early planetesimals or embryos. Among the key questions that need to be addressed with gravitoturbulent planetesimal formation is: 1) what are the size of bound clumps of dust and 2) will these bound clumps collapse into solid objects or be disrupted by high gas or particle velocities? 

In this paper, we use 3D local hydrodynamical simulations to model the interaction of particles in a gravitationally unstable but non-fragmenting disk. We include particle backreaction and self-gravity for a self-consistent treatment of particle dynamics. In Sections \ref{sec:gravitationalcollapse} and \ref{sec:model} we detail the necessary theory of dust dynamics in marginally gravitationally unstable disks and the numerics of the \textsc{Pencil} code\footnote{http://pencil-code.nordita.org/}, respectively. Section \ref{sec:analysis} details the identification of clumps and the scaling relations to determine clump masses. In Section \ref{sec:results} we discuss the results, focusing on particle and gas velocities and the masses of the resulting particle clumps. We continue in Section \ref{sec:discussion} with a discussion of the implications on protoplanetary disk evolution and planet formation and summarize with our main conclusions in Section \ref{sec:conclusion}.

\section{Gravitational Collapse}
\label{sec:gravitationalcollapse}

When the density of a region is high enough in a diffuse medium, the high density region can undergo gravitational collapse. Whereas the collapse of an adiabatic gas is resisted by the thermal pressure of the gas \citep{Jeans1902,Gammie2001,Kratter2011}, this thermal pressure does not provide dust particles any extra stability against collapse. Instead, the random motion of the particles determines the resistance to gravitational collapse. Collapse may be prevented on scales closer to the final solid object due to the terminal velocity of dust collapsing in a gas \citep{WahlbergJansson2017,Visser2021}, but that is beyond the resolution capabilities of our simulations. Here, we only consider that the collapse of dust in a self-gravitating disk relies on concentrating enough dust locally to overcome internal turbulent diffusion processes which may arise through coupling to turbulent motions of the gas \citep{Klahr2020,Klahr2021}. Without considering any particle movement, a cloud of particles at orbital radius $R$ around a star of mass $M_{*}$ should be unstable to collapse when the cloud has a density higher than the Hill density
\begin{equation} \label{eq:hilldensity}
\rho_{\mathrm{Hill}} = \frac{9}{4\pi} \frac{M_{*}}{R^{3}}.
\end{equation}
This should not be confused with the Roche density of a body \citep{Chandrasekhar1963,Shi2013},
\begin{equation} \label{eq:rochedensity}
\rho_{R} = 3.5 \frac{M_{*}}{R^{3}},
\end{equation}
which is the minimum density a body needs to remain bound. In this paper, the former is used to establish the collapse criterion of diffusive particles in a disk and the latter is used to identify persistent dense objects.

Turbulent gas flows can simultaneously facilitate and disrupt the local concentration of dust. Turbulent motions can distribute dust between eddies or to local gas density maxima \citep{Squires1991,Cuzzi2001,Johansen2006a}, however at the same time impart diffusive particle motions, keeping particles from remaining at high densities for long. The strength of this dust diffusion depends on how well particles couple to the gas turbulent motions \citep{Youdin2007a} and is defined in one direction by the diffusion constant $D$ as the average distance a particle moves from its original position per unit time \citep[i.e.][]{Johansen2007,Yang2009}:
\begin{equation} \label{eq:diffusionconstant}
D_{d,x} \equiv \frac{1}{2}\frac{d\langle | x(t) - x(0) |^{2} \rangle}{dt}\,,
\end{equation}
where $x(t)$ is the position of a particle at time $t$.

Turbulence in gravitoturbulent disks is largely subsonic, but regions of supersonic turbulence can be found at various heights above the disk midplane \citep{Shi2014}, particularly near the midplane of spiral density features \citep{Cossins2009,Riols2020}. The motion of larger dust particles is largely determined by the gravitational interaction with the gaseous spirals, leading to particle concentration within the spirals. For a dust particle, the gravitational force from the spiral is stronger than the aerodynamic drag force when \citep{Shi2016,Baehr2021}
\begin{equation} \label{eq:duststructure}
\frac{Q}{\mathrm{St}} \lesssim 1\,,
\end{equation}
where $\mathrm{St}$ is the particle's Stokes number, normally proportional to the particle's size.
For a marginally stable disk $Q \approx 1$, this condition is met for $\mathrm{St} \gtrsim 1$. Dust particles settle more efficiently in gravitoturbulent disks than in disks which do not include self-gravity, due to the gas self-gravity and the anisotropic turbulence that has weaker vertical particle diffusion compared to radial transport \citep{Riols2020,Baehr2021a}. Including the drift of particles to density maxima, this results in a dense particle layer predominantly at the spiral midplane.

Smaller particles which are well-coupled to the gas are more affected by the aerodynamic drag force than the gravitational force from the gas, which leads to lower velocity dispersions \citep{Booth2016}. The velocity dispersion $\sigma$ for a collection of particles is defined as
\begin{equation} \label{eq:dispersiondefinition}
\sigma = \sqrt{\frac{1}{N_{\mathrm{par}}} \sum_{j}^{N_{\mathrm{par}}} |\bm{w}_{j} - \langle \bm{w}\rangle |^{2}},
\end{equation}
where $\bm{w}_{j}$ is the velocity of particle $j$ and $\langle \bm{w}\rangle$ is the average velocity of all $N_{\mathrm{par}}$ particles in a clump or in the whole simulation domain. As with particle diffusion, greater dispersion of the particle velocities makes it more difficult for particles to gravitationally collapse.
\begin{figure*}[t]
\centering
\includegraphics[width=0.5\textwidth]{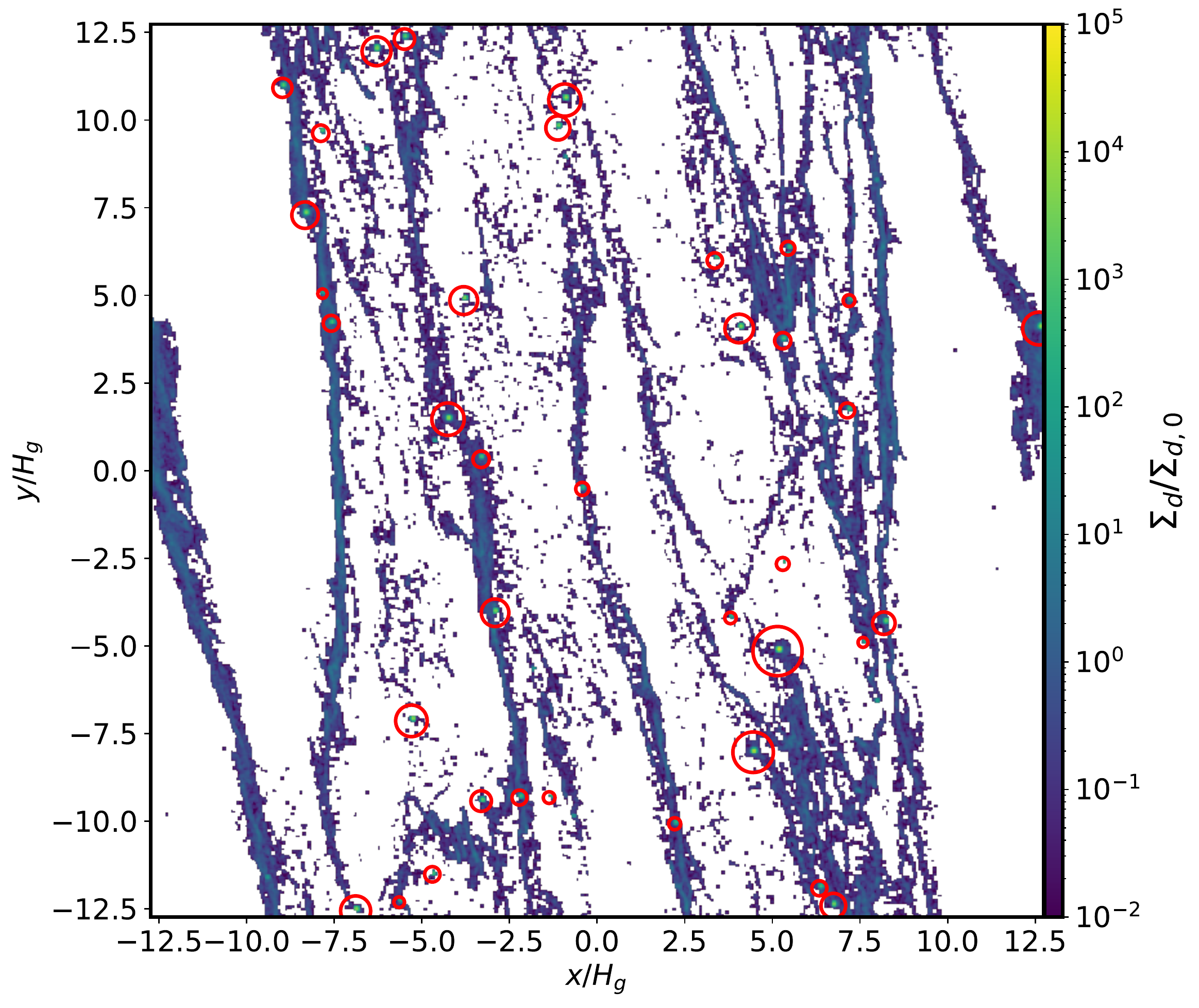}%
\includegraphics[width=0.5\textwidth]{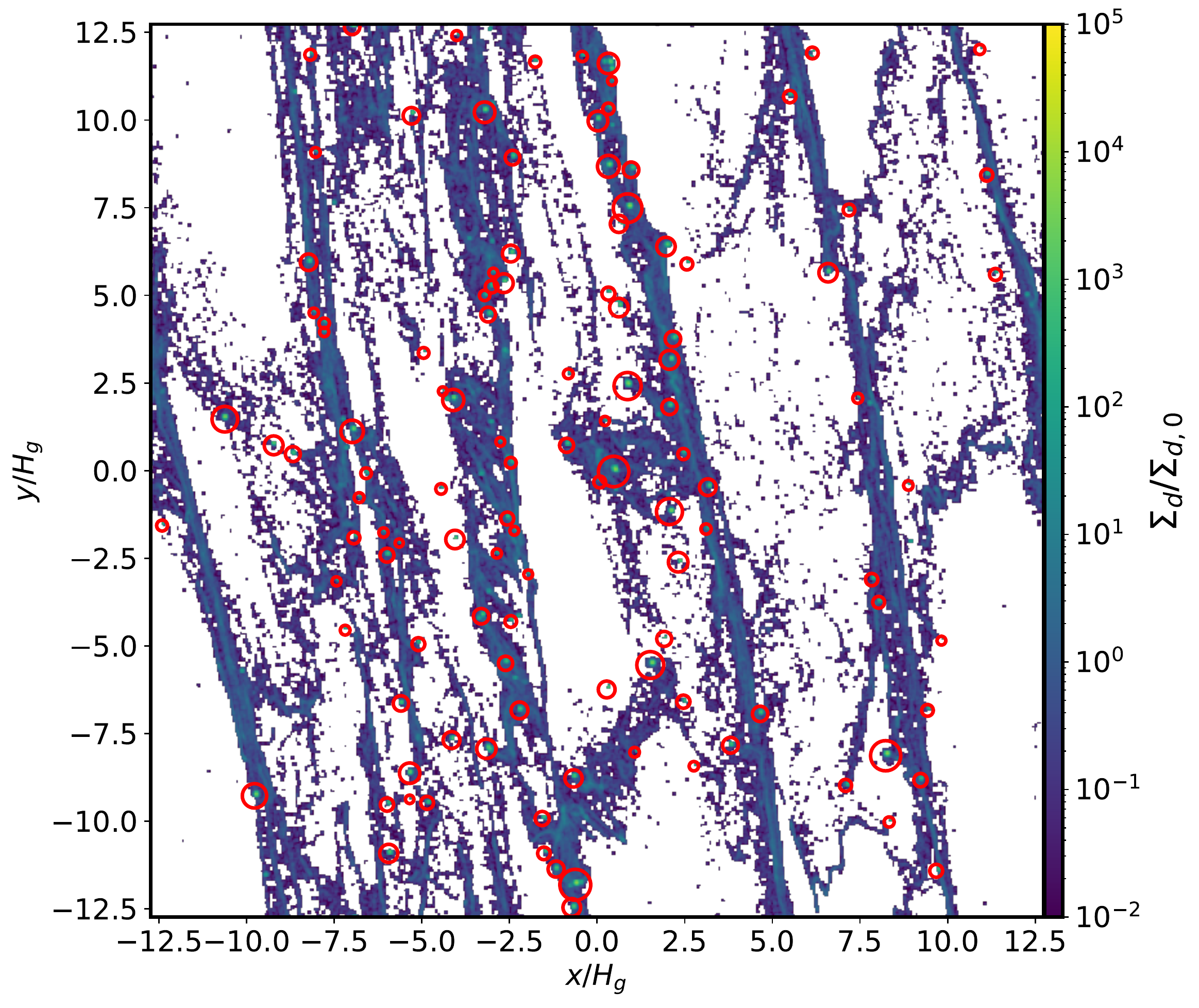}
\caption{The vertically-integrated dust particle density $\Sigma_{d}$ at $t=60\,\Omega^{-1}$ for simulations without dust backreaction (left) and with dust backreaction (right), normalized by the initial dust particle surface density $\Sigma_{d,0} = 0.01\Sigma_{g,0}$. All particles are size $\mathrm{St}=1$, which migrate efficiently towards regions of high gas density, triggering collapse of high density particle clouds. Red circles indicate locations where the particle density at one or more cells is above Roche density \eqref{eq:rochedensity} and gravitationally bound. The size of the circle indicates the Hill radius \eqref{eq:hillradius} of each dust clump.}
\label{fig:gravitoturb}
\end{figure*}
Including particle backreaction changes the picture slightly. In this case, a collection of particles can push the gas around as they move through the gas, but this effect is small when local dust-to-gas mass ratios are less than one. Only when the dust strongly accumulates does it begin to affect gas velocities.

When one includes diffusive movement of particles, the condition for particle cloud collapse can be derived by balancing the gravitational force and the particle diffusion. A cloud of particles with uniform density $\rho_{\mathrm{d}}$ and size $r$ can be prevented from collapsing at large scales by tidal shear and at small scales by the turbulent diffusion generated by a particle-gas instability, such as the streaming instability \citep[SI;][]{Johansen2007,Yang2021} or the Kelvin-Helmholtz instability \citep[KHI;][]{Weidenschilling1980}. A stability parameter $Q_{\mathrm{d}}$ is derived in \citet{Klahr2021,Gerbig2020}:
\begin{equation} \label{eq:gerbigparameter}
Q_{\mathrm{d}} = \frac{3}{2}\frac{Q}{\epsilon Z}\sqrt{\frac{\delta_x}{\mathrm{St}}} < 1.
\end{equation}
In other words, the gas gravitational stability parameter $Q$ is modified by dimensionless radial diffusion coefficient $\delta_x = D_{d,x}/(c_s H_g)$, the particle's Stokes number $\mathrm{St}$, the overall dust-to-gas mass ratio of the clump $Z$ and gas scale height $H_g$. The factor $\epsilon$ is a measure of the local enhancement of the dust surface density as a result of radial-azimuthal turbulent concentration of the dust $\epsilon \equiv \Sigma_{d,max}/\langle \Sigma_d \rangle$.
\begin{figure*}[t]
\centering
\includegraphics[width=0.5\textwidth]{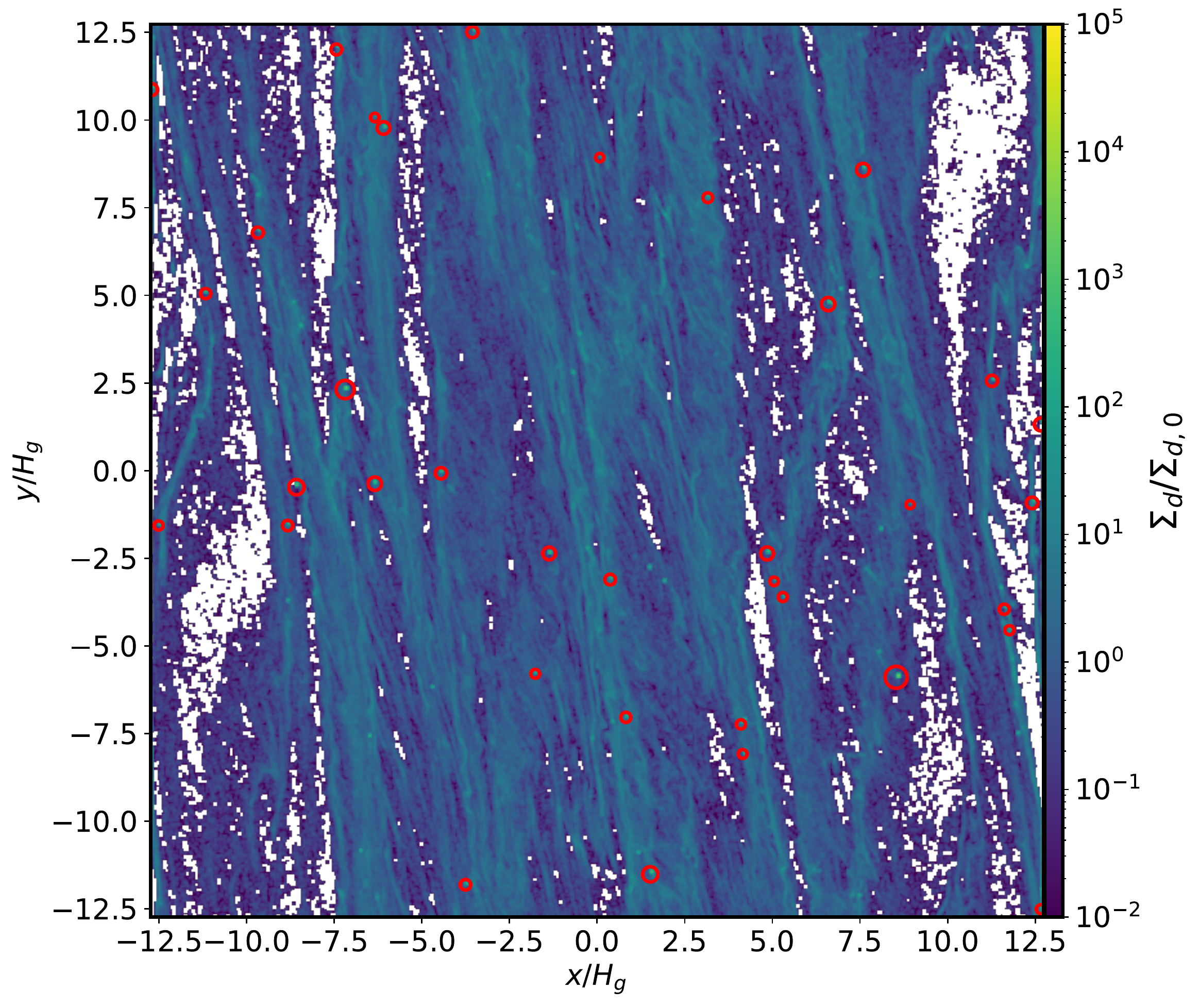}%
\includegraphics[width=0.5\textwidth]{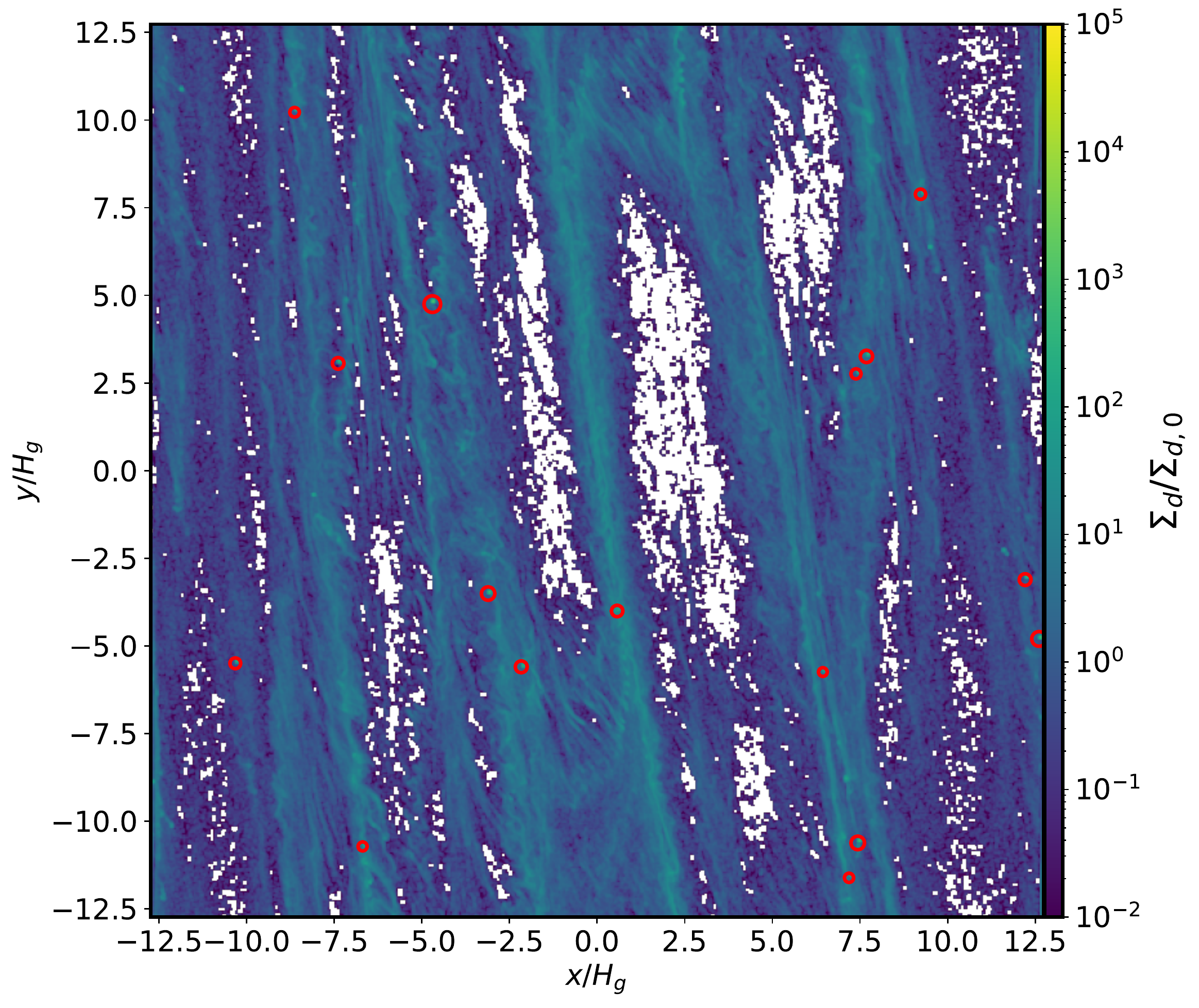}
\caption{Simulations similar to those in Figure \ref{fig:gravitoturb} but with $\mathrm{St}=0.1$. The panel on the left is the simulation without dust backreaction and the panel on the right includes backreaction. Since dust of this size does not drift towards the gas structures as efficiently, clumps are fewer and less massive, but still of considerable mass. }
\label{fig:gravitoturb2}
\end{figure*}
In the simulations of this study, we test the above criterion for scales where GI turbulence dominates dust diffusion and SI and KHI are not resolved. Furthermore, no radial pressure gradient is included in the simulations of this paper, so even with sufficient resolution SI does not develop.

\section{Model}
\label{sec:model}

\begin{figure*}[t]
\centering
\includegraphics[width=0.5\textwidth]{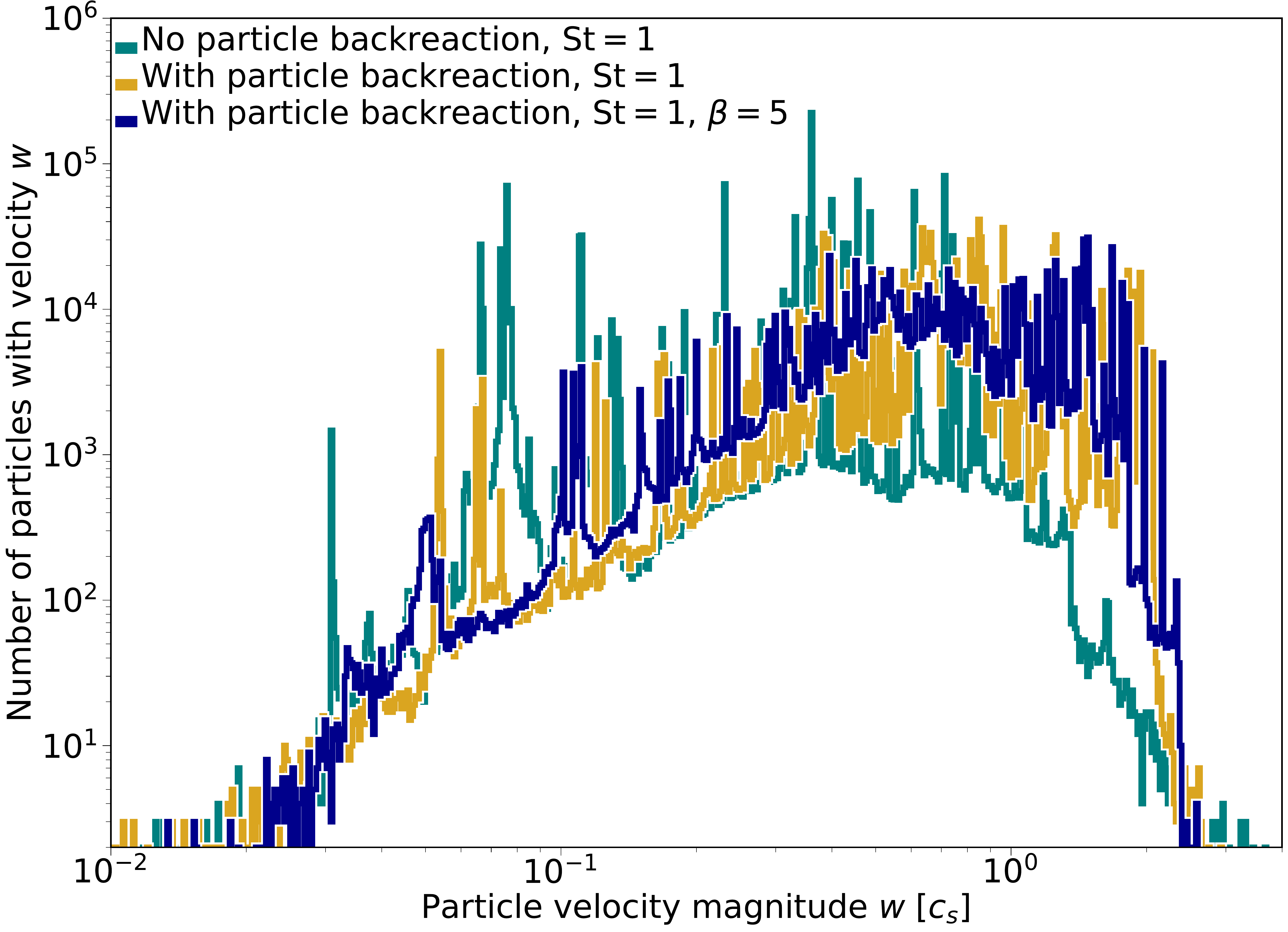}%
\includegraphics[width=0.5\textwidth]{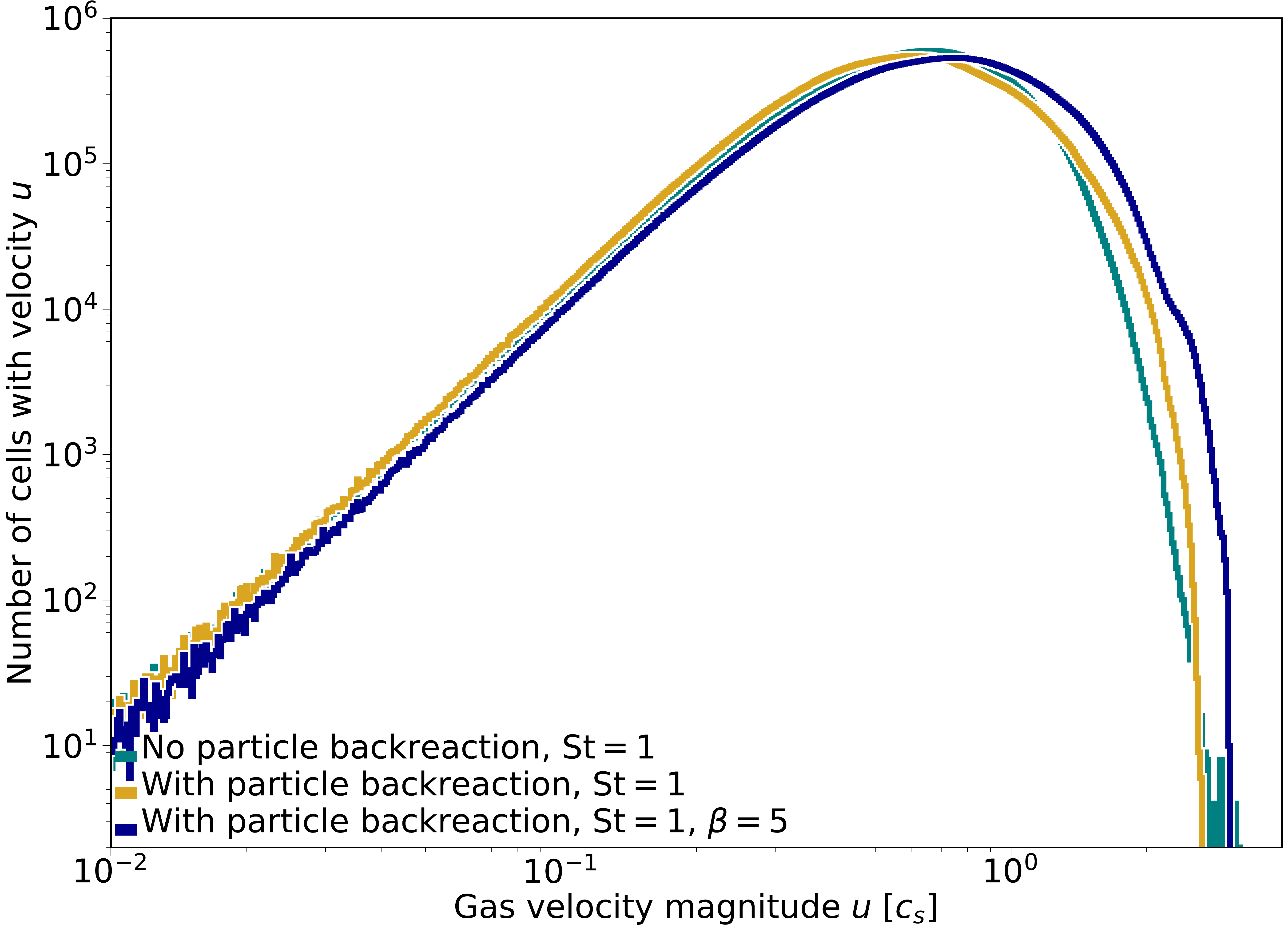}
\includegraphics[width=0.5\textwidth]{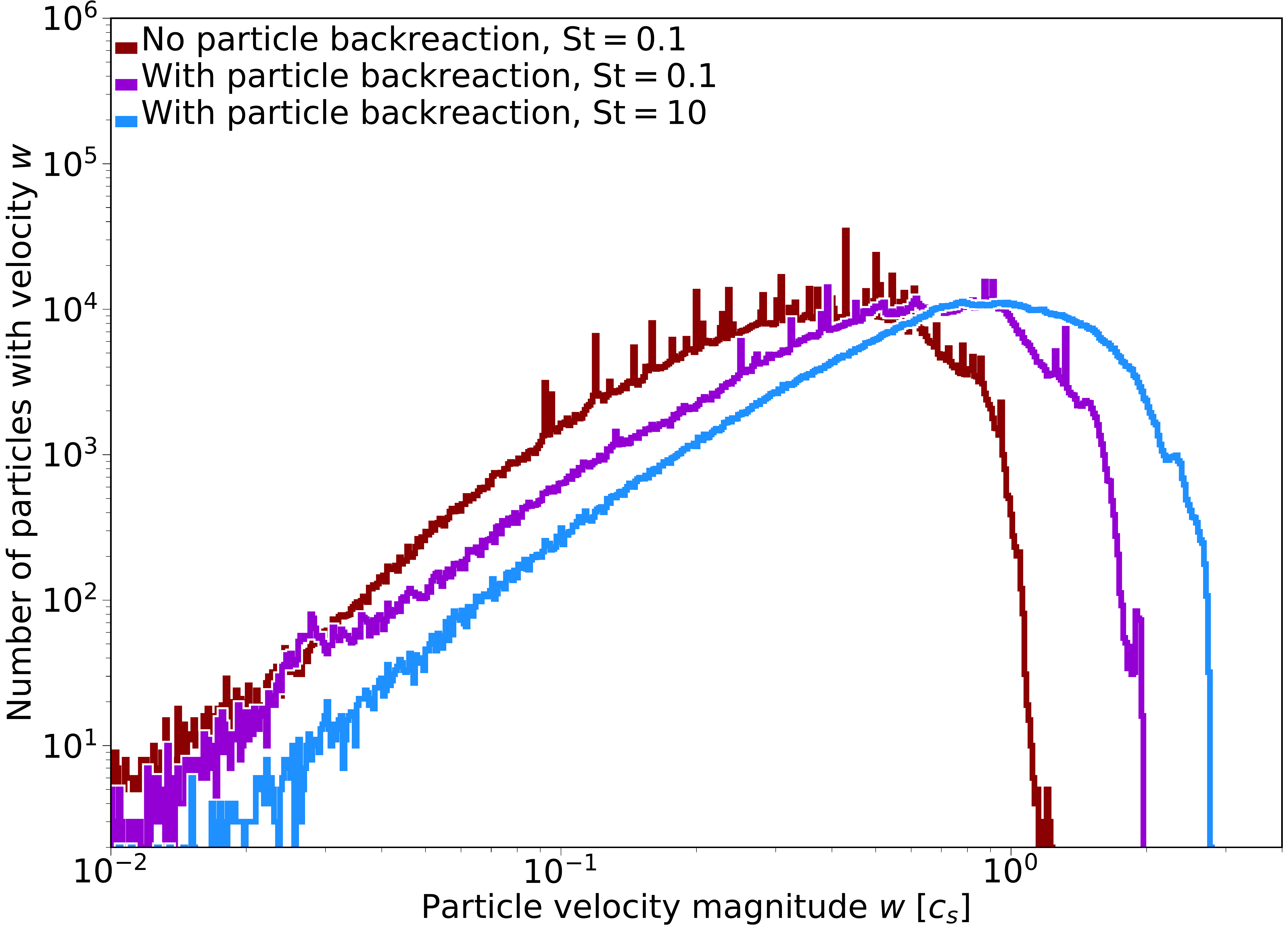}%
\includegraphics[width=0.5\textwidth]{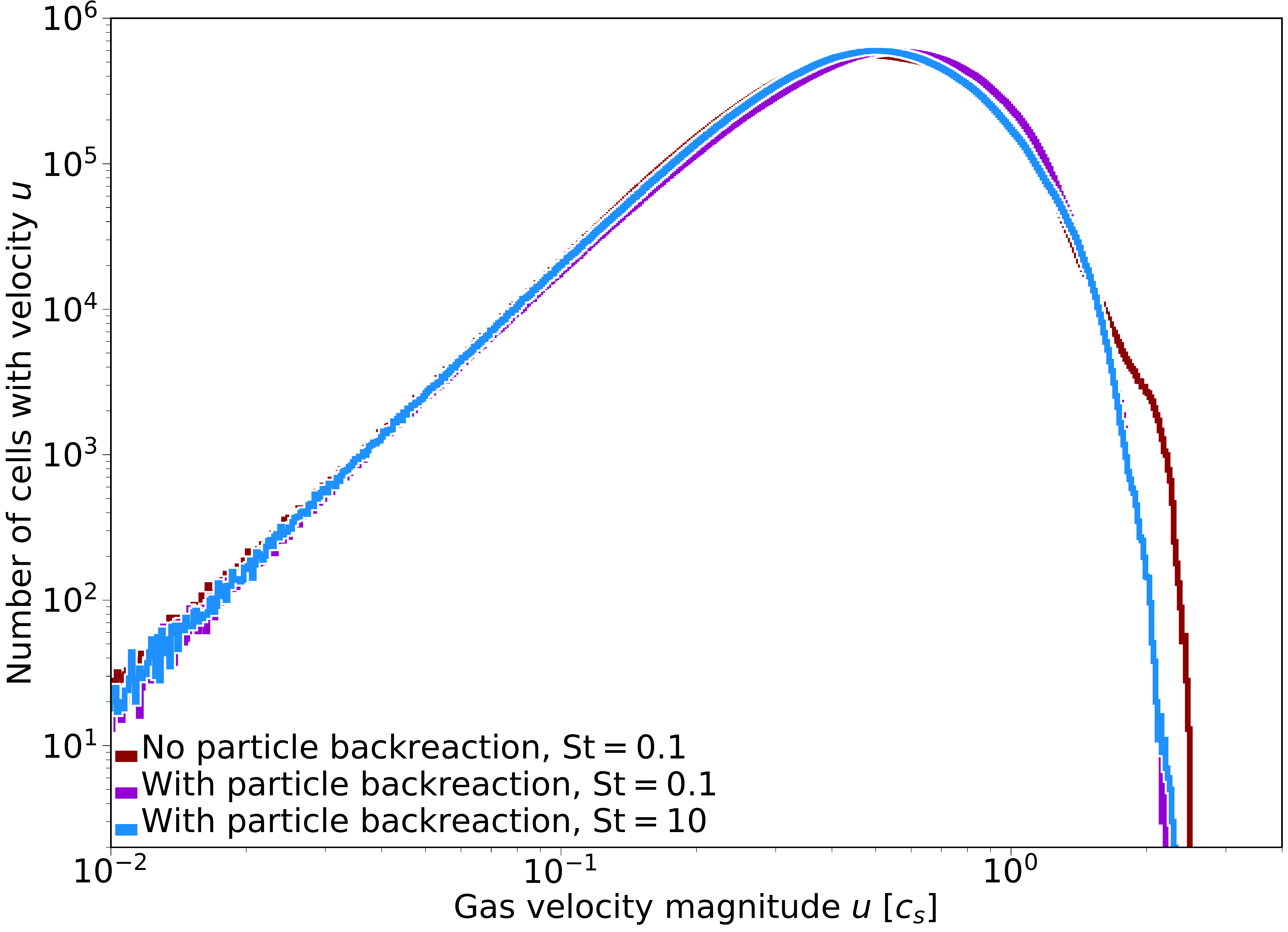}
\caption{Top row: On the left, the histogram of particle velocity magnitudes in simulations with $\mathrm{St}=1$ particles at $t=60\,\Omega^{-1}$. Sharp peaks in the velocity space correspond to concentrations of particles that have low relative velocities (within the clump), but may have a high collective motion. These peaks thus show the velocity of the center of mass of each particle clump. On the right, the histogram of gas velocities at the same point in time. Bottom row: particle velocity (left) and gas velocity (right) histograms of simulations with other particle sizes $\mathrm{St}=0.1$ and $\mathrm{St}=10$.}
\label{fig:gasdustvelocities}
\end{figure*}

For this study we conduct 3D hydrodynamic shearing box simulations of a self-gravitating disk with Lagrangian super-particles embedded in the Eulerian mesh using the \textsc{Pencil} code \citep{Brandenburg2003,PencilCode2021}. Both gas and particles are treated as self-gravitating and particle backreaction is calculated on the gas by mapping the change of particle momentum due to the gas drag back to the grid with triangular-shaped clouds \citep{Youdin2007}. Local simulations that allow the Toomre wavelength $\sim 2\pi H_g$ to be well-resolved in the radial and azimuthal coordinates ($x$ and $y$ in the linearized coordinates, respectively) should avoid spurious fragmentation \citep{Truelove1997,Nelson2006}. All simulations use $512^{2}\times 256$ grid cells with box lengths $L_{x}=L_{y}=(80/\pi) H_g$ and $L_{z}=(40/\pi) H_g$ such that $\Delta x=\Delta y =\Delta z \simeq 0.05\, H_g$, where $H_g$ is the vertical scale height of the gas at the initial uniform temperature. At this grid resolution, we run the simulations with 512 processors up to a simulation time of $t = 80\,\Omega^{-1}$. This takes around 100-200 hours ($\sim$50,000 to 100,000 cpu-hours per simulation) depending on the particle size, where smaller particle sizes require smaller timesteps to resolve high dust-to-gas ratios. We run a high resolution simulation with $\Delta x=\Delta y =\Delta z \simeq 0.025\, H_g$ up to $t = 30\,\Omega^{-1}$ when an initial distribution of clumps has formed to check for convergence.

Shearing box simulations use hydrodynamic equations which are linearized and transformed into Cartesian coordinates co-rotating in a Keplerian disk, where $q = -d\mathrm{ln}\Omega / d\mathrm{ln}R = 3/2$ is the shear parameter:
\begin{figure*}[t]
\centering
\includegraphics[width=0.5\textwidth]{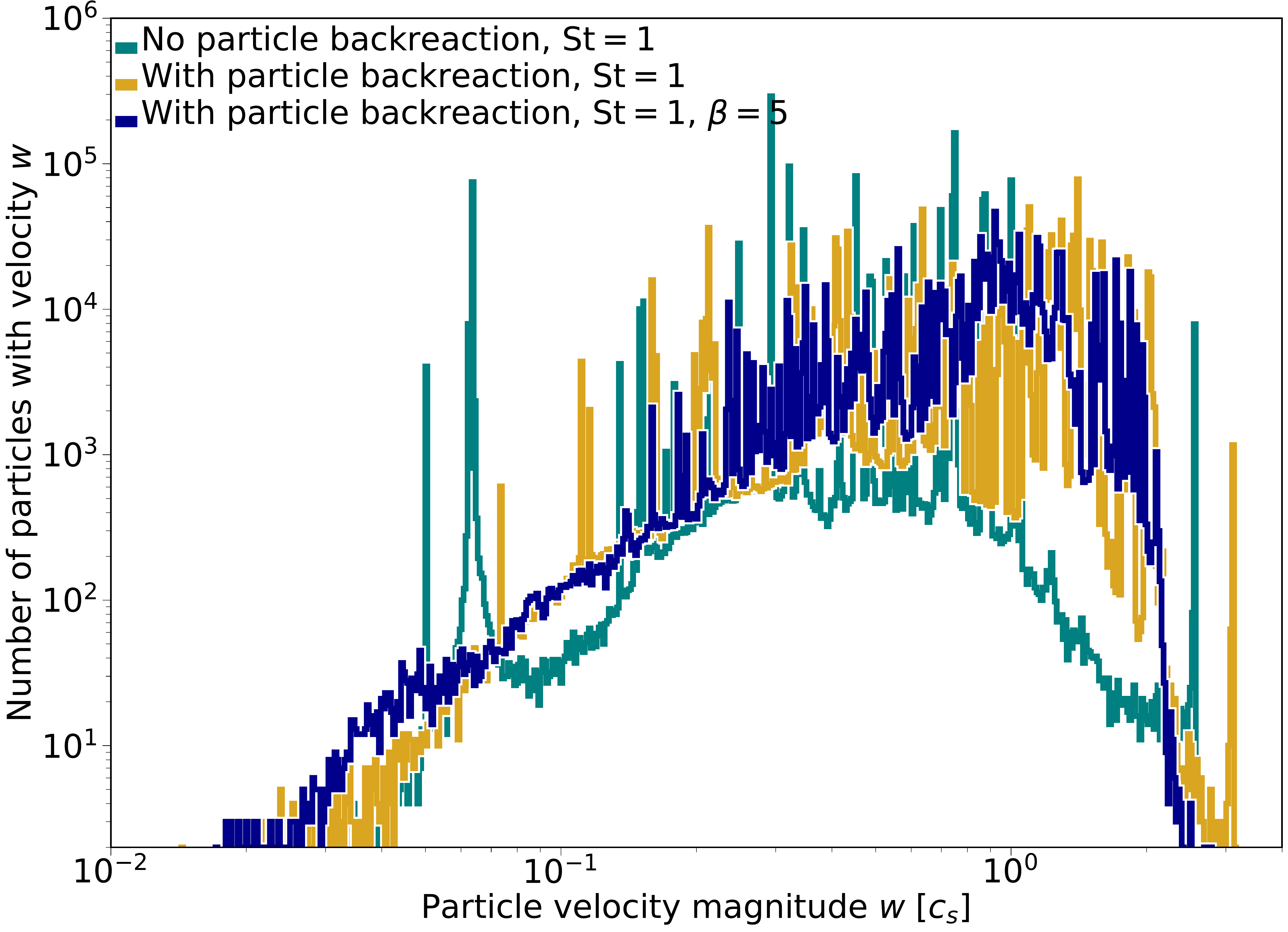}%
\includegraphics[width=0.5\textwidth]{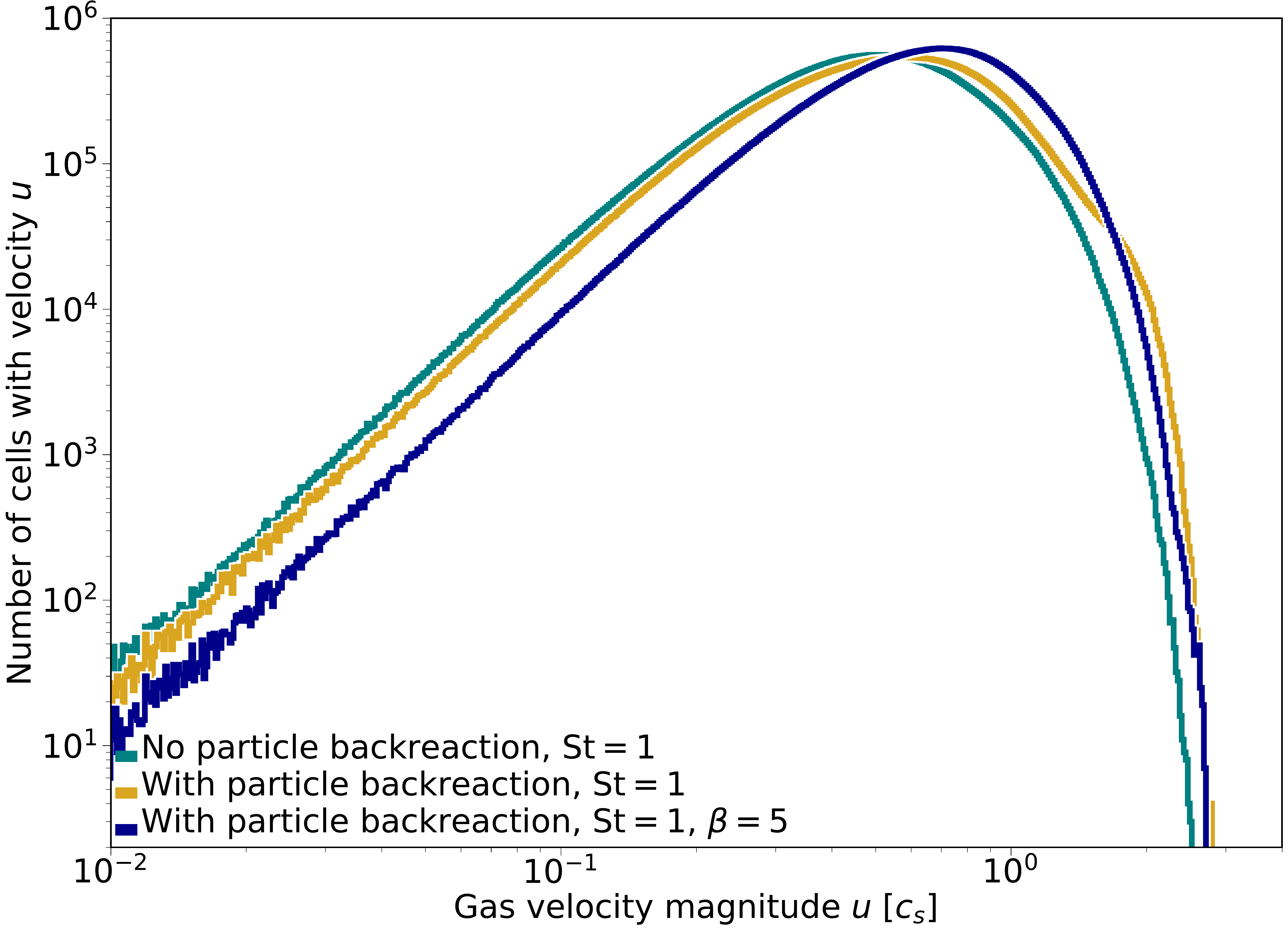}
\includegraphics[width=0.5\textwidth]{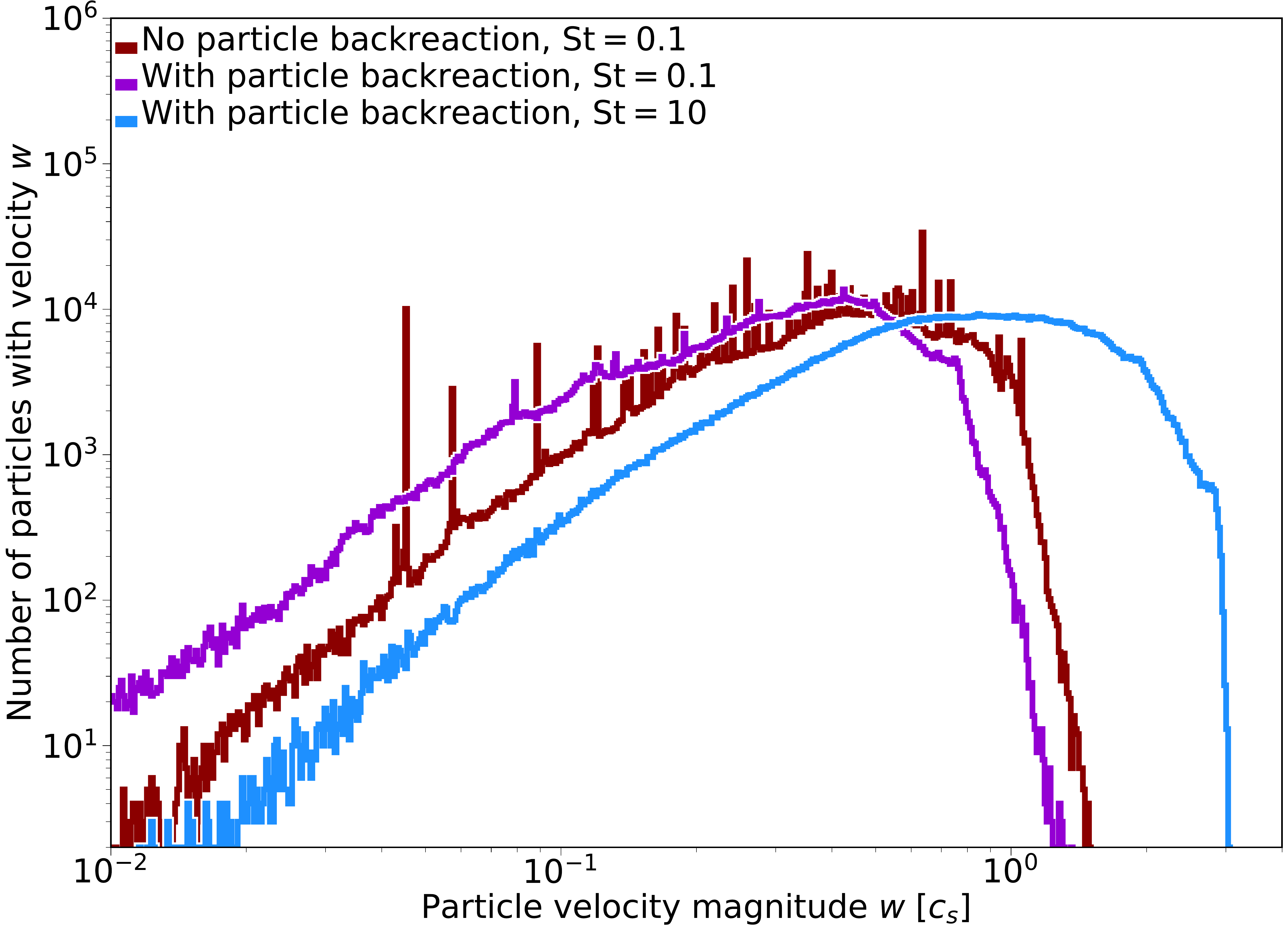}%
\includegraphics[width=0.5\textwidth]{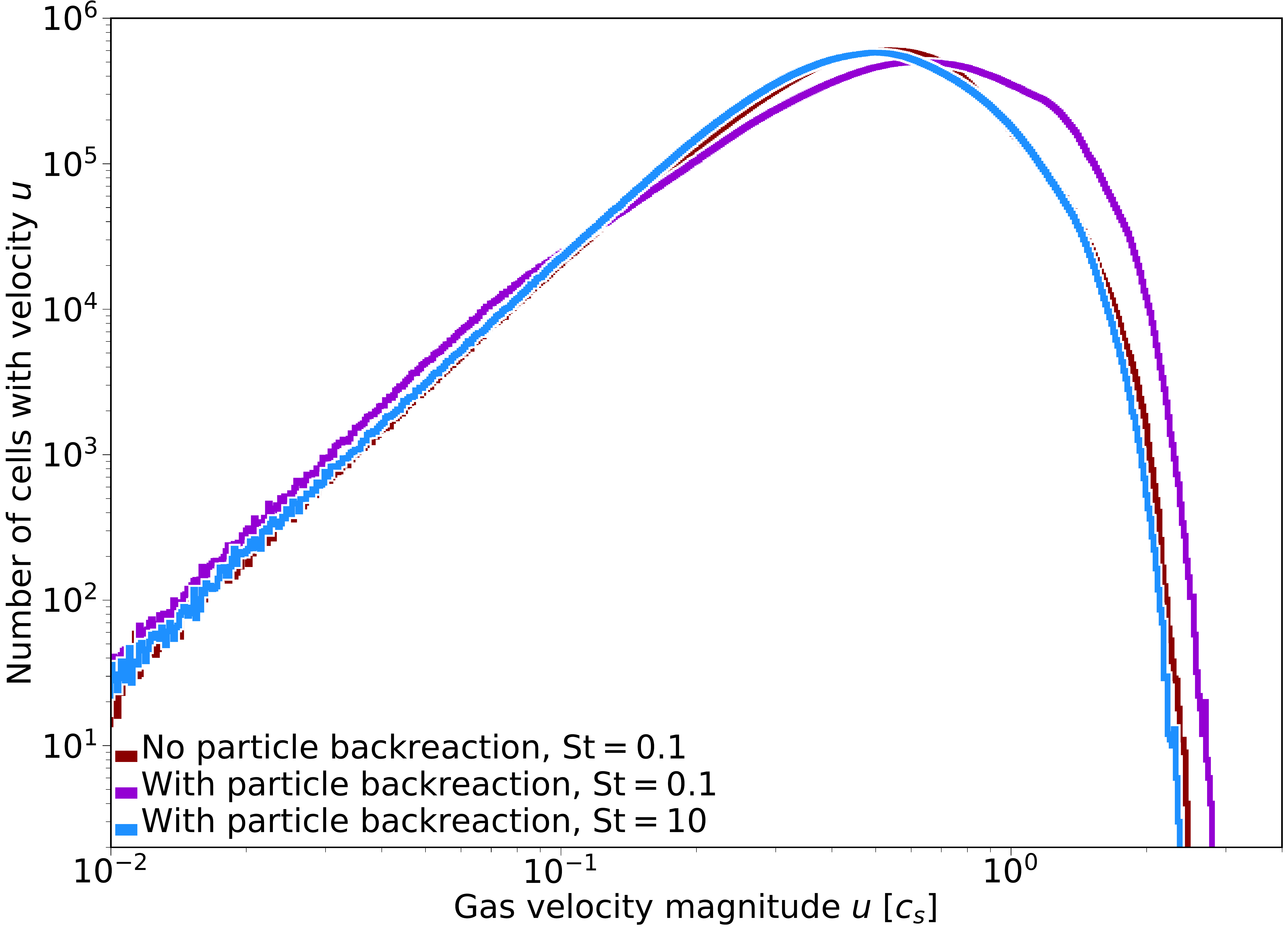}
\caption{Same as Figure \ref{fig:gasdustvelocities}, but at $t = 72\,\Omega^{-1}$. The top panels show the simulations with particle size $\mathrm{St}=1$, and have the same general features as the earlier snapshot. Both simulations with particle backreaction have more dust and gas at trans- to supersonic velocities, compared to the simulations without backreaction.}
\label{fig:gasdustvelocities2}
\end{figure*}
\begin{align}
\frac{\partial {\rho_{\mathrm{g}}}}{\partial t} &- q\Omega x\frac{\partial {\rho_{\mathrm{g}}}}{\partial y} + \nabla\cdot(\rho_{\mathrm{g}}\bm{u}) = f_{D}(\rho_{\mathrm{g}}) \label{eq:finalmassconserve} \\
\frac{\partial \bm{u}}{\partial t} &- q\Omega x\frac{\partial \bm{u}}{\partial y} + \bm{u}\cdot\nabla\bm{u} = -\frac{\nabla p}{\rho_{\mathrm{g}}} + q\Omega u_{x}\bm{\hat{y}} \nonumber \\ 
& - 2\bm{\Omega}\times\bm{u} - \nabla\Phi - \mathbf{g} - \frac{\varepsilon}{\tau_{s}} ( \bm{u} - \bm{w} ) + f_{\nu}(\bm{u}) \label{eq:finalmomconserve} \\
\frac{\partial s}{\partial t} &-q\Omega x\frac{\partial s}{\partial y} + (\bm{u} \cdot \nabla)s = -\frac{\Lambda}{\rho_{\mathrm{g}} T} + f_{\chi}(s). \label{eq:finalenergyconserve}
\end{align}
In equations \eqref{eq:finalmassconserve} - \eqref{eq:finalenergyconserve}, $\mathbf{u}$ is the gas velocity deviation from the background shear velocity in the local box, $\bm{w}$ is the particle velocity which imparts a backreaction onto the gas proportional to the local dust-to-gas ratio $\varepsilon \equiv \rho_\mathrm{d} / \rho_\mathrm{g}$, $\rho_{\mathrm{g}}$ is the gas density, $\rho_{\mathrm{d}}$ is the dust density.
The thermodynamic variable is the specific entropy $s$, while $p$ is the gas pressure and $T$ is the gas temperature. The vertical gravitational acceleration $\bm{g} = g\hat{\bm{z}}$ due to the central potential is a linear profile modified with zero acceleration near the $z$-boundary to avoid an abrupt discontinuity at the periodic vertical boundary.

We use an ideal equation of state such that
\begin{equation} \label{eq:eos}
p = (\gamma - 1)\rho_g e,
\end{equation}
where $\gamma \equiv c_p / c_v \equiv 5/3$ is the adiabatic index, $c_p$ and $c_v$ are the specific heat capacities at constant pressure and volume, respectively, and $e = c_v T$ is the specific internal energy. Heat can be generated through the dissipation of shocks via a dimensionless shock viscosity of $\nu_{sh} = 5.0$ \citep[see][]{Lyra2008a} and through compression of the gas. The relationship between the gas temperature $T$, the sound speed $c_s$, and entropy is given by
\begin{equation}
c_{s}^2 = (\gamma - 1) c_p T = c_{s,0}^2 \exp [\gamma s/c_p + (\gamma - 1)\ln (\rho/\rho_0)]
\end{equation}
where $c_{s,0}$ is the initial uniform speed of sound. Heat is lost via the simple $\beta$-cooling prescription
\begin{equation} \label{eq:coolingfunction} 
\Lambda = \frac{\rho_g (c_{\textnormal{s}}^{2} - c_{\textnormal{s,irr}}^{2})}{(\gamma -1) t_{\textnormal{c}}} 
\end{equation}
with the cooling timescale $t_{\mathrm{c}}$ parametrized as $t_{\mathrm{c}} = \beta\Omega^{-1}$. We include a background irradiation term which is different from most previous simulations. This background irradiation is important at the outer regions of protoplanetary disks where GI prevails. The background temperature mimics the effect of stellar irradiation \citep{DAlessio1998}, keeping the gas from dropping below the initial $Q_0$ due to a low local gas temperature. We choose $c_{\textnormal{s,irr}} = c_{s,0}$ such that the disk cools towards the initial uniform temperature. This cooling prescription has no dependence on local variations in optical depth and thus all regions cool with the same timescale. With a more realistic treatment of thermodynamics, the opacity is dominated by small dust grains and an increase of the particle density can increase the local cooling timescale.

The term $\epsilon ( \bm{u} - \bm{w} ) / \tau_{s}$ in Equation \eqref{eq:finalmomconserve} is the backreaction of particles with stopping time $\tau_{s}$ (see Equation~\eqref{eq:epsteinregime} below), included or removed depending on whether this effect is active or inactive. Hyperdissipation is applied with the terms $f_{D}(\rho_{\mathrm{g}})$, $f_{\nu}(\bm{u})$, $f_{\chi}(s)$ which for each has the form
\begin{equation} \label{eq:hyperdiff}
f(\xi) = \nu_3(\nabla^{6}\xi),
\end{equation}
with hyperdissipation constant $\nu_3 = 2.5\, H_g^{6}\Omega$, which leads to a mesh Reynolds number of about 0.15 \citep{Yang2012,Lyra2017} during fully developed gravitoturbulent state.

The gravitational potential of the gas and dust is solved in Fourier space by transforming the density to find the potential at wavenumber $k$ and transforming the solution back into real space. The solution is shifted in the  $y$-direction by applying a phase in the Fourier space such that the shear periodic boundary conditions are accounted for \citep[i.e.][]{Johansen2007a}. The solution to the Poisson equation in Fourier space at sheared wavenumber $k$ is
\begin{equation} \label{eq:gravpotential}
\Phi(\bm{k}, t) = -\frac{4\pi G\rho(\bm{k}, t)}{\bm{k}^2},
\end{equation}
where the density of the particles and gas is combined $\rho = \rho_{\mathrm{g}} + \rho_{\mathrm{d}}$ to produce a potential $\Phi = \Phi_{\mathrm{g}} + \Phi_{\mathrm{d}}$. This means that self-gravity includes both gas and particles and affects both components.
\begin{figure*}[t]
\centering
\includegraphics[width=0.5\textwidth]{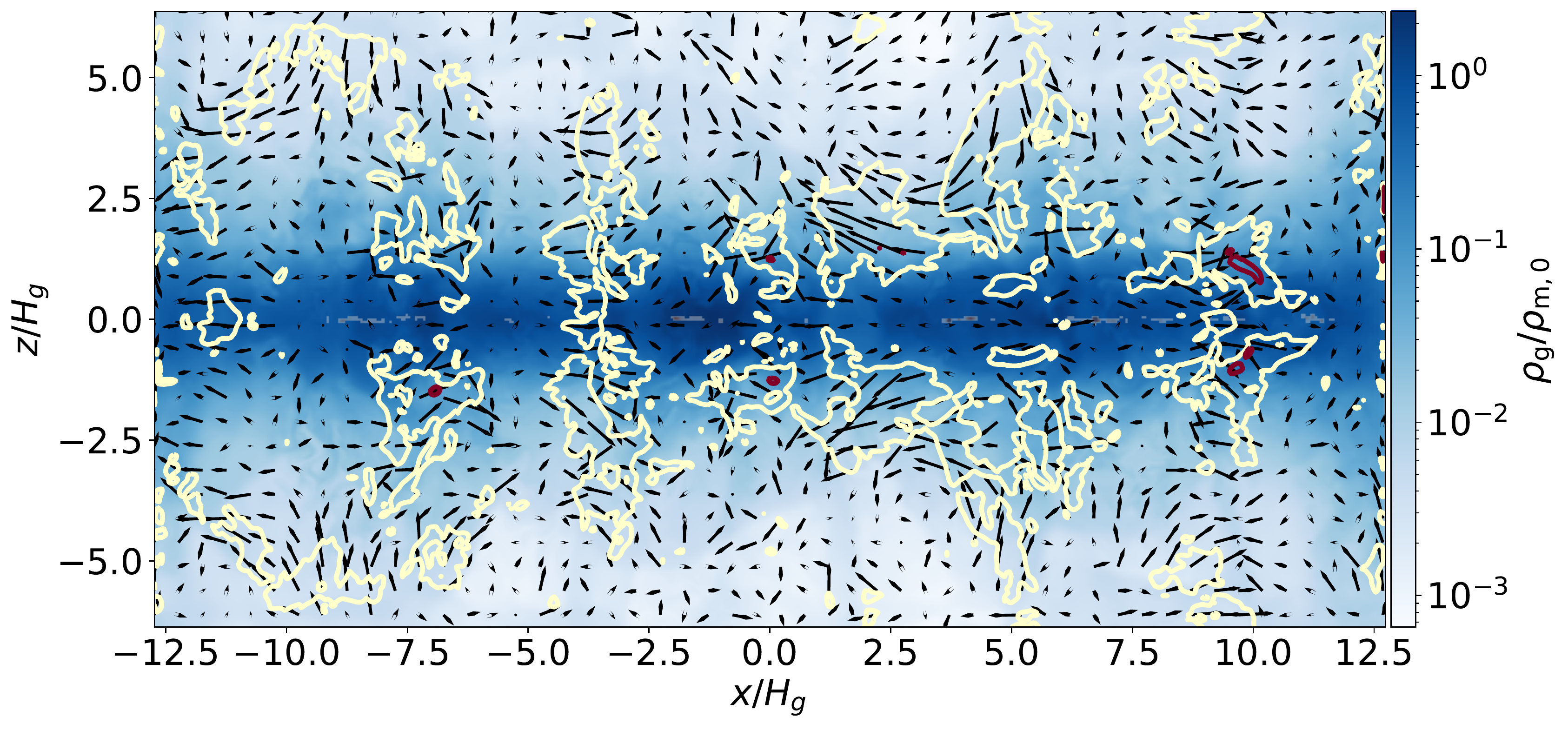}%
\includegraphics[width=0.5\textwidth]{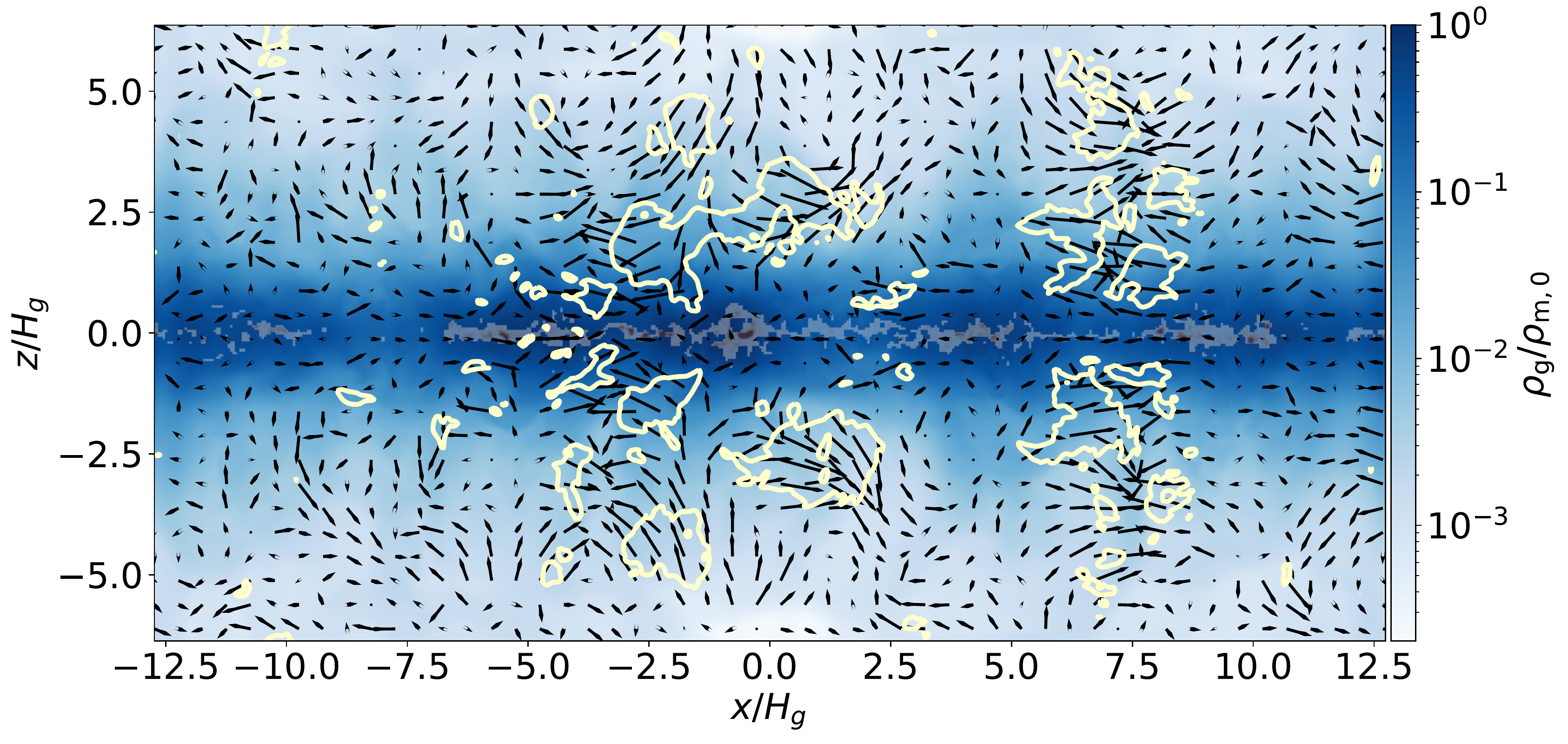}
\caption{A pair of $x$-$z$ slices at $t = 60\Omega^{-1}$ through simulations with $\mathrm{St}=1$ dust particles  (left) and $\mathrm{St}=0.1$ dust particles (right), both without dust backreaction. Vectors indicate the gas flow direction and magnitude in the $x-z$ plane, while contours in yellow and purple indicate regions where the 3D gas velocity magnitude is above Mach number equals 1 and 2, respectively. Blue colors show the gas density structure and gray particles plotted as triangular-shaped clouds, which are settled close to $z=0$.}
\label{fig:supersonicflows}
\end{figure*}

Every particle in our simulations, also known as a superparticle, represents a collection of solids such that the $i$-th superparticle has position $\bm{x}^{(i)}$ and velocity $\bm{w}^{(i)}$ as in \citet{Youdin2007,Yang2016}. It follows that
\begin{align}
\frac{d \bm{w}^{(i)}}{d t} &= -2\Omega \times \bm{w}^{(i)} +q\Omega w^{(i)}_x\bm{\hat{y}} - \nabla\Phi\\
&- \frac{1}{\tau_{s}} \left( \bm{w}^{(i)} - \bm{u}(\bm{x}^{(i)}) \right), \label{eq:particlevel}\\
\frac{d \bm{x}^{(i)}}{d t} &= \bm{w}^{(i)} - q\Omega x^{(i)} \bm{\hat{y}}. \label{eq:particlepos}
\end{align}
In the Epstein regime, the stopping time is proportional to the particle size as \citep{Weidenschilling1977}
\begin{equation} \label{eq:epsteinregime}
\tau_{\mathrm{s}} = \frac{a\rho_{\bigcdot}}{c_{\mathrm{s}}\rho_{\mathrm{g}}}
\end{equation}
where $a$ is the particle diameter and $\rho_{\bigcdot}$ is the material density of an individual dust particle. The dimensionless friction time $\tau_{f}$ (also referred to as the Stokes number $\mathrm{St}$) is the particle stopping time normalized by the orbital time $\Omega^{-1}$
\begin{equation} \label{eq:stokenum}
\mathrm{St} = \tau_{f} = \tau_{\mathrm{s}}\Omega.
\end{equation}
Larger particles have higher Stokes numbers, are less coupled to small scale gas motions, and retain their initial motion for longer. As a corollary, smaller particles with low Stokes numbers are well-coupled and closely move with the gas. Particles are added such that the initial distribution maintains a physically motivated metallicity of $Z=0.01$, roughly that of the interstellar medium (ISM). The vertical dust distribution follows a vertical Gaussian profile with the same width as the gas. Particle mass is calculated from the total gas mass and the specified metallicity (see also Section~\ref{sec:analysis}).

Since the radial pressure gradient is set as zero, particles do not drift in the disk. The contribution of particles to the gravitational potential is initially minuscule and the potential is dominated by the gas distribution, but with rapid dust settling to the midplane, the entire dust layer is covered by 5 to 10 grid cells ($\sim 0.25 - 0.5\, H_g$) \citep{Baehr2021a}.  Since our simulations include $1.5\times 10^{6}$ particles we resolve the midplane layer ($\sim 5$ cells thick) with roughly $1.1$ particles per cell.

\section{Analysis}
\label{sec:analysis}

Our simulation parameters are summarized in Table \ref{tab:sims}. Simulations without dust backreaction are indicated by the `noBR' segment in the name and `BR' if dust backreaction is included. Three sizes of particles are used, denoted by `S' ($\mathrm{St=0.1}$), `L' ($\mathrm{St=1}$) and `XL' ($\mathrm{St=10}$). A final simulation at a higher resolution is indicated by `HR'. Included in Table \ref{tab:sims} are time and space averaged diagnostics of the gas gravitational stability $Q$, Reynolds and gravitational stresses $\alpha_R$ and $\alpha_G$, defined as
\begin{equation}\label{eq:reynoldsstress}
\alpha_{R} = \frac{2}{3} \frac{\langle \rho u_{x}u_{y} \rangle}{\langle \rho c_{\mathrm{s}}^{2} \rangle},
\end{equation}
and
\begin{equation}
\alpha_{G} = \frac{2}{3} \frac{\langle g_{x}g_{y} \rangle}{4\pi G\langle \rho c_{\mathrm{s}}^{2} \rangle},
\end{equation}
where $g_{x}$ and $g_{y}$ are the gravitational accelerations in the radial and azimuthal directions, respectively. As in \citet{Baehr2021a}, these stresses are calculated as simple volume averages over the time range $t = 50\,\Omega^{-1}$ to $t = 80\,\Omega^{-1}$ when the gas gravitoturbulence has been established. The sum of these stresses yields the total $\alpha$ stress in the disk \citep{Shakura1973,Gammie2001}.

Also measured in Table~\ref{tab:sims} are the dimensionless particle diffusion constants $\delta \equiv D_d / c_s H_g$ and velocity dispersions $\sigma$ in the vertical and radial directions. Particle diffusions are calculated in \textsc{Pencil} using the scheme from \citet{Yang2009} where the particle displacement $x(t) - x(0)$ of all particles is averaged as in Equation \eqref{eq:diffusionconstant}. The particle velocity dispersion is calculated using all particles according to Equation \eqref{eq:dispersiondefinition} and averaged in time between $t = 50\,\Omega^{-1}$ and $t = 80\,\Omega^{-1}$.
While diffusion may sometimes be assumed equal in each direction, turbulent processes in protoplanetary disks can be non-isotropic \citep{Zhu2015b} and may result in particle concentration in certain structures \citep{Yang2018}.

We identify dense clumps based on the cells in which the dust density is at or above Roche surface density 
\begin{equation}
\Sigma_{R} \approx \sqrt{2\pi}H_g\rho_{R} = 8.8 \frac{M_{*} H_g}{R^{3}} = 8.8\frac{\Omega^{2}H_g}{G}.
\end{equation}
Since the dust scale height $H_d$ is much smaller than the gas scale height $H_g$, this serves as a criterion to identify the largest dust clouds, although smaller bound clouds may exist. The Roche surface density may not be a definitive criterion since a dense particle layer may be susceptible to disruption from shockwaves that can transverse the particle concentration faster than it can collapse \citep{Shi2013}. Nevertheless, we find the Roche density to be a suitable threshold for the identification of persistent clumps.

We search for clumps with a $3\times 3$ grid cell footprint over the vertically integrated particle density. The particles are so settled to the midplane that the volumetric density in the midplane and surface density are closely correlated. Within each $3\times 3$ patch, cells which are at or above the Roche surface density are used to calculate the total dust mass of each clump $M_{c}$. From the clump mass, we compute its Hill radius $R_{H}$, the region around the clump where its gravity dominates the tidal force. Most clumps are small enough that the Hill radius is within the $3\times 3$ footprint, but for larger clumps, the mass estimate might neglect some particles that are within the Hill radius, but outside the $3\times 3$ grid cells, where particle density is usually low compared to the central density (Figures~\ref{fig:gravitoturb} and~\ref{fig:gravitoturb2}).
\begin{equation} \label{eq:hilllimit}
R_{H} = \left( \frac{M_{c}}{3M_{*}} R^{3} \right)^{1/3}
\end{equation}
We can rewrite $M_{*}/R^{3}$ from our adopted Toomre stability parameter $Q_{0} = 1.02$ and initial midplane density $\rho_{m,0}$ as
\begin{equation}
\frac{M_{*}}{R^{3}} = \frac{\Omega^{2}}{G} = \sqrt{2\pi}\pi \rho_{m,0} Q_{0}
\end{equation}
which yields an expression for the Hill radius of a clump
\begin{equation} \label{eq:hillradius}
R_{H} = \left( \frac{M_{c}}{3\sqrt{2\pi}\pi \rho_{m,0} Q_{0}} \right)^{1/3} \approx \left( \frac{M_{c}}{24 \rho_{m,0}}. \right)^{1/3}
\end{equation}
We plot the Hill radius of each identified bound clump in Figures \ref{fig:gravitoturb} and \ref{fig:gravitoturb2} which range from two to eight grid cells in radius.

\begin{figure*}[t]
\centering
\includegraphics[width=0.85\textwidth]{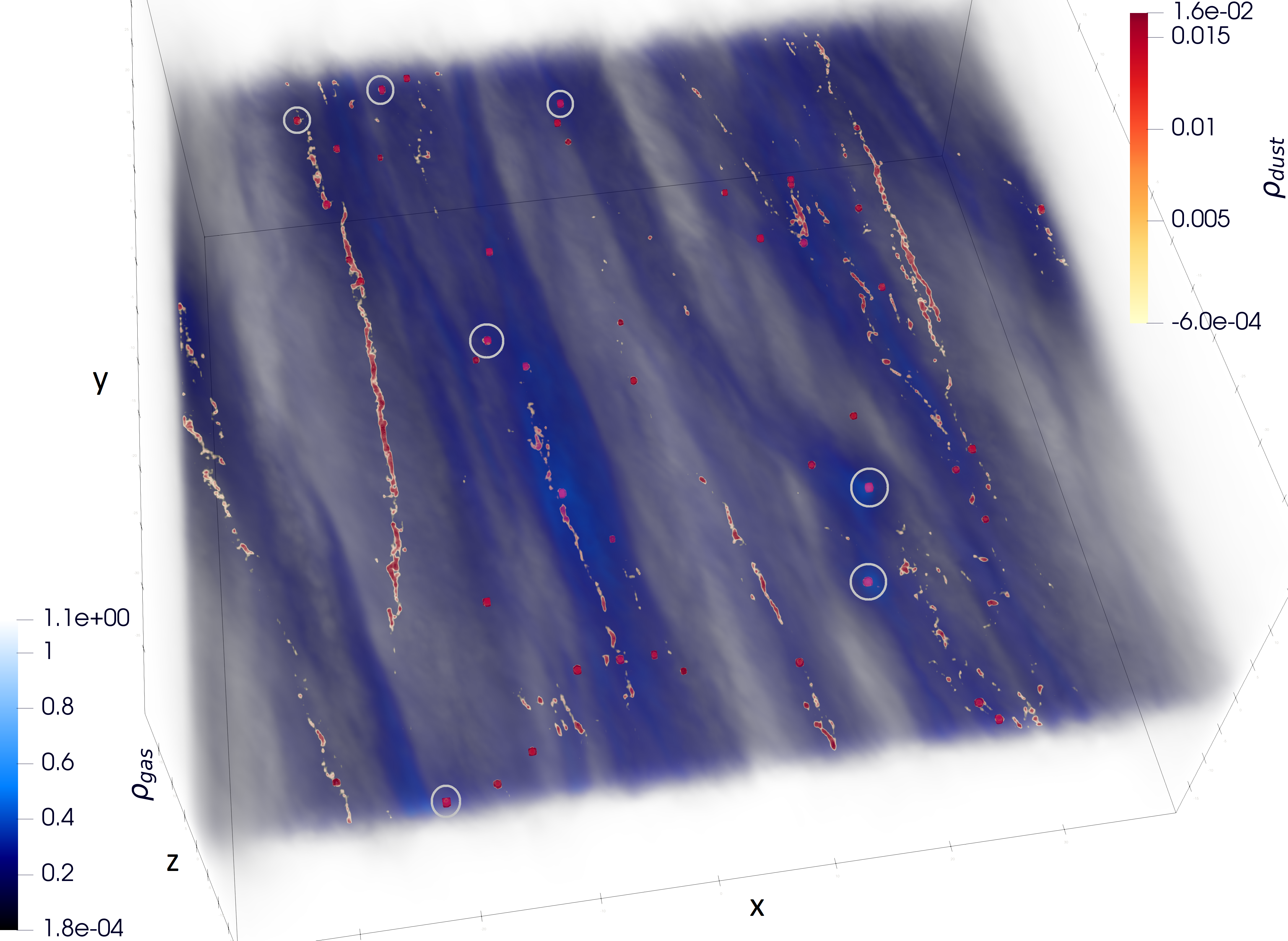}
\caption{A 3D render of the snapshot at the left in Figure \ref{fig:gravitoturb}. Dense particle clumps show as bright red regions while dense particle streams that have not collapsed are orange to yellow. The most massive clouds $> 1 M_{\oplus}$ are circled. Values for gas and particle densities are normalized by the initial mid-plane gas density.}
\label{fig:3drender}
\end{figure*}

Similar to \citet{Schafer2017}, we provide the following scaling relations, using a mean molecular weight of $\mu = 2.33$ at a reference distance of $50\, au$ around a $1\, M_{\odot}$ star, to convert our results in physical units for the sound speed $c_s$, orbital frequency $\Omega$, orbital period $P$, and gas scale height $H_{g}$.
\begin{equation}
c_s = 199 \left( \frac{T}{11.25\,K} \right)^{1/2} \, m\, s^{-1}
\end{equation}
\begin{equation}
\Omega = 1.78 \times 10^{-2} \left( \frac{R}{50\, au} \right)^{-3/2} \left( \frac{M_{*}}{1 M_{\odot}} \right)^{1/2}\, yr^{-1}
\end{equation}
\begin{equation}
P = 353 \left( \frac{R}{50\, au} \right)^{3/2} \left( \frac{M_{*}}{1 M_{\odot}} \right)^{-1/2}\, yr
\end{equation}
\begin{equation}
H_{g} = 2.36 \left( \frac{T}{11.25\,K} \right)^{1/2} \left( \frac{R}{50\, au} \right)^{3/2} \left( \frac{M_{*}}{1 M_{\odot}} \right)^{-1/2}\, au.
\end{equation}
The scale height changes as the local temperature fluctuates, so the $H_g$ we define here and use throughout the paper is based on the initial uniform temperature, i.e., with the speed of sound at $c_s = c_{s,\mathrm{irr}} = c_{s,0}$. The above definitions mean the initial surface density is
\begin{equation}
\begin{split}
\Sigma_{0} &= \frac{c_{s,0}\Omega}{\pi G Q_{0}} \\
&= 53\left( \frac{T}{11.25\,K} \right)^{1/2} \left( \frac{R}{50\, au} \right)^{-3/2} \left( \frac{M_{*}}{1 M_{\odot}} \right)^{1/2}\,g\,cm^{-2}.
\end{split}
\end{equation}
\begin{figure*}[t]
\centering
\includegraphics[width=0.95\textwidth]{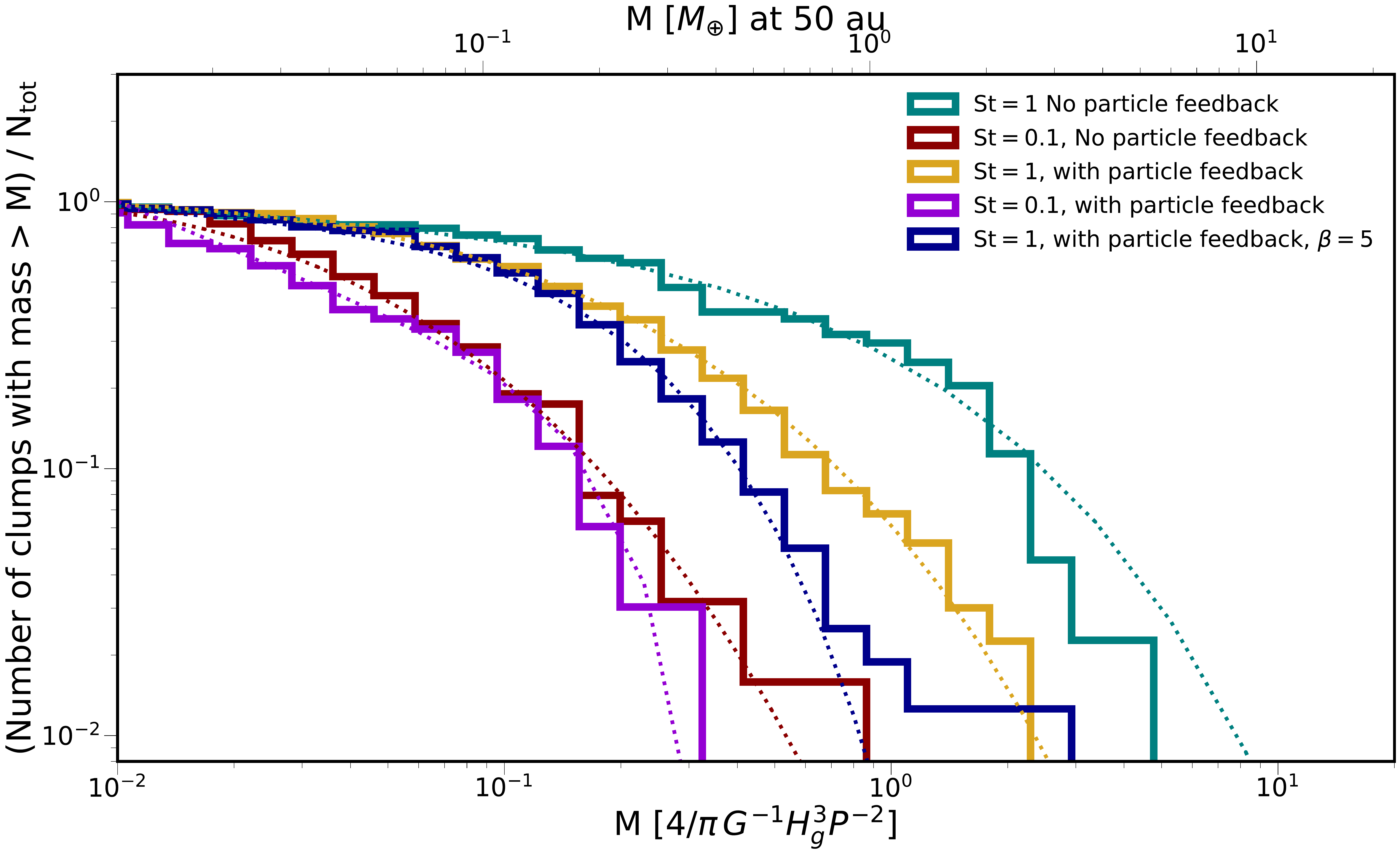}%
\caption{Normalized cumulative mass distribution of particle concentrations in the five simulations where particles reach Roche density and are gravitationally bound. Plotted distributions are at a time when there is little variation in subsequent snapshots and is considered stable. No particle clumps were identified in the simulation using particles of size $\mathrm{St}=10$. Mass is displayed in code units (bottom axis), physical units assuming the shearing box is at 50 au (top axis; see Equation \eqref{eq:codemassunit}) and fit with an exponentially-tapered power law (dotted curves).}
\label{fig:cumulativemassdistribution}
\end{figure*}
In our models, we adopt a length of $H_g = \pi$, a unit time of $\Omega^{-1} = P / 2\pi$, and $G=1$, so our unit volume density is
\begin{equation}
\begin{split}
\hat{\rho}_{0} &= \frac{\Omega^{2}}{G}\\
&= 4.7 \times 10^{-12} \left( \frac{R}{50\, au} \right)^{-3} \left( \frac{M_{*}}{1 M_{\odot}} \right)\, g\, cm^{-3},
\end{split}
\end{equation}
and our unit mass is
\begin{equation} \label{eq:codemassunit}
\begin{split}
\hat{M}_{0} &= \frac{4}{\pi}\frac{H_{g}^{3}}{GP^{2}}\\
&=6.8 \times 10^{27} \left( \frac{T}{11.25\,K} \right)^{3/2} \left( \frac{R}{50\, au} \right)^{3/2}\\ 
&\times\left( \frac{M_{*}}{1 M_{\odot}} \right)^{-1/2}\, g = 1.1\,M_{\oplus}
\end{split}
\end{equation}
With this value for code unit of mass, we derive the total gas mass in the entire simulation box (assuming at values from above at 50 au)
\begin{equation}
\begin{split}
M_{\mathrm{total,gas}} &= \Sigma_{0} L_{x}L_{y} = 80^{2} \hat{M}_{0}/Q_0 \\
&= 4.3 \times 10^{31}\, g = 0.021\,M_{\odot}.
\end{split}
\end{equation}
With this total gas mass, and the dust mass assumed is to be at a 1:100 ratio, the total dust mass is
\begin{equation}
\begin{split}
M_{\mathrm{total,dust}} &= 64 \hat{M}_{0}/Q_0 \\
&= 4.3 \times 10^{29}\, g = 72\,M_{\oplus}.
\end{split}
\end{equation}
Dividing it by the number of super particles ($1.5 \times 10^{6}$) yields the mass per super particle
\begin{equation}
\begin{split}
M_{\mathrm{sp}} = 2.8 \times 10^{23}\, g = 4.8 \times 10^{-5}\,M_{\oplus}.
\end{split}
\end{equation}
Using the above scaling relations, we can translate the code units to physical units at any disk radius of interest. The surface density we adopt results in a disk to star mass ratio around 0.1, which is appropriate for self-gravitating disks and within the constraints of observations, e.g. GM Aurigae \citep{Schwarz2021}. Additionally, many protoplanetary disks could be optically thick in submillimeter observations, resulting in a larger disk mass than estimates based on the optically thin assumption \citep{Zhu2019a}. In the next section, we proceed with the simulation results, starting with the dust and gas velocities, and how they depend on our choice of particle size and dust backreaction, followed by the mass of the dust concentrations that form.

\begin{figure*}[t]
\centering
\includegraphics[width=0.98\textwidth]{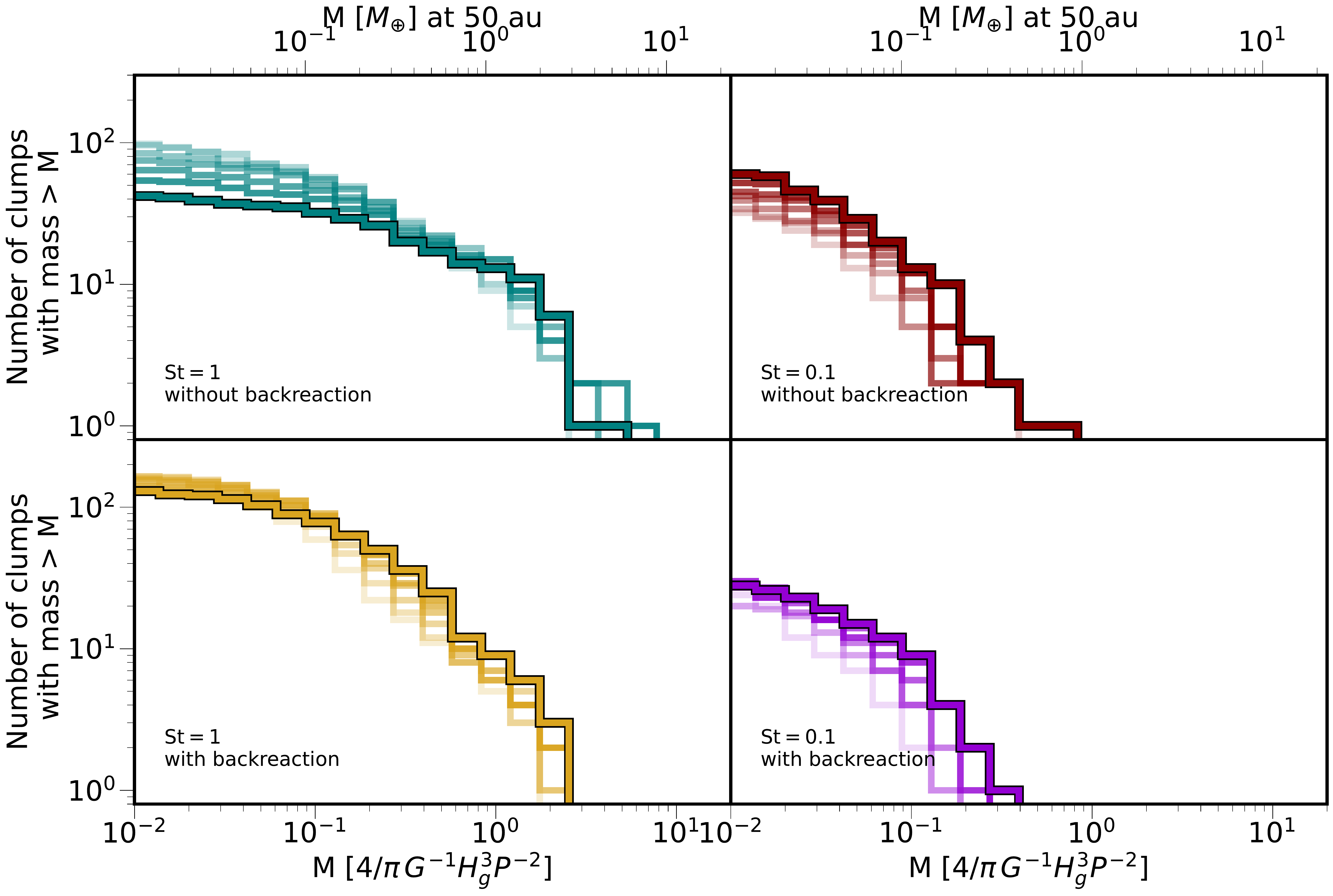}%
\caption{Evolution of the clump mass distribution in select simulations. The progression from lighter to darker colors indicates the progression of time in increments of $4 \Omega^{-1}$ until the snapshot which represents the converged distribution which is indicated with a black outline. Colors in each panel correspond the simulations of the same color in Figure \ref{fig:cumulativemassdistribution}.}
\label{fig:cumulativemassdistributionevolution}
\end{figure*}

\section{Results}
\label{sec:results}

Our simulations show that particles clump into dense, bound clouds up to a few Earth masses if placed at $\sim$50\,au. This is exemplified in Figure \ref{fig:gravitoturb}, which shows the locations where particle densities within a single grid cell are above Roche surface density\footnote{A movie of the evolution of these clumps can be found at \url{https://youtu.be/hqtwMIe1dwk}}. The red circles drawn around them indicate the Hill radius of each clump of particles, representing the approximate region around the clump where the gravity of the clump is stronger than the stellar gravity. Figure \ref{fig:gravitoturb2} shows similar clumping behavior for dust with a smaller Stokes number (St=0.1), although clumps are smaller and fewer.

\subsection{Gas and Particle Velocities}
\label{subsec:velocityresults}

In the top two panels of Figure \ref{fig:gasdustvelocities}, we compare the dust velocities in simulations with $\mathrm{St}=1$ particles and the gas velocities of all cells in the simulation domain. Particles of this size drift towards pressure maxima with the greatest efficiency. The numerous spikes in the velocity histogram are where a significant number of particles have the same velocity. Thus, they roughly represent the center-of-mass velocity of each clump.  For the case without particle backreaction, numerous large clumps form and most remain together at subsonic speeds. 

Dust backreaction, combined with the high local enhancements in dust-to-gas ratio brought on by the clumping pushes the gas velocity in a number of cells near the mid-plane further into the supersonic regime. Increasing the strength of the GI turbulence by decreasing the cooling time $\beta$ has a similar effect and shifts even more gas towards supersonic speeds. Both effects increase the number of particles in the supersonic regime, but also accelerate some clumps. Overall, most particle clumps are still within two times the sound speed. Figure \ref{fig:gasdustvelocities2} shows the same simulations at a later time $\Delta t = 12 \,\Omega^{-1}$. The simulations in the top panel of both Figures \ref{fig:gasdustvelocities} and \ref{fig:gasdustvelocities2} with $\mathrm{St}=1$ have similar velocity patterns in gas and particle velocities.

\begin{figure*}[t]
\centering
\includegraphics[width=0.85\textwidth]{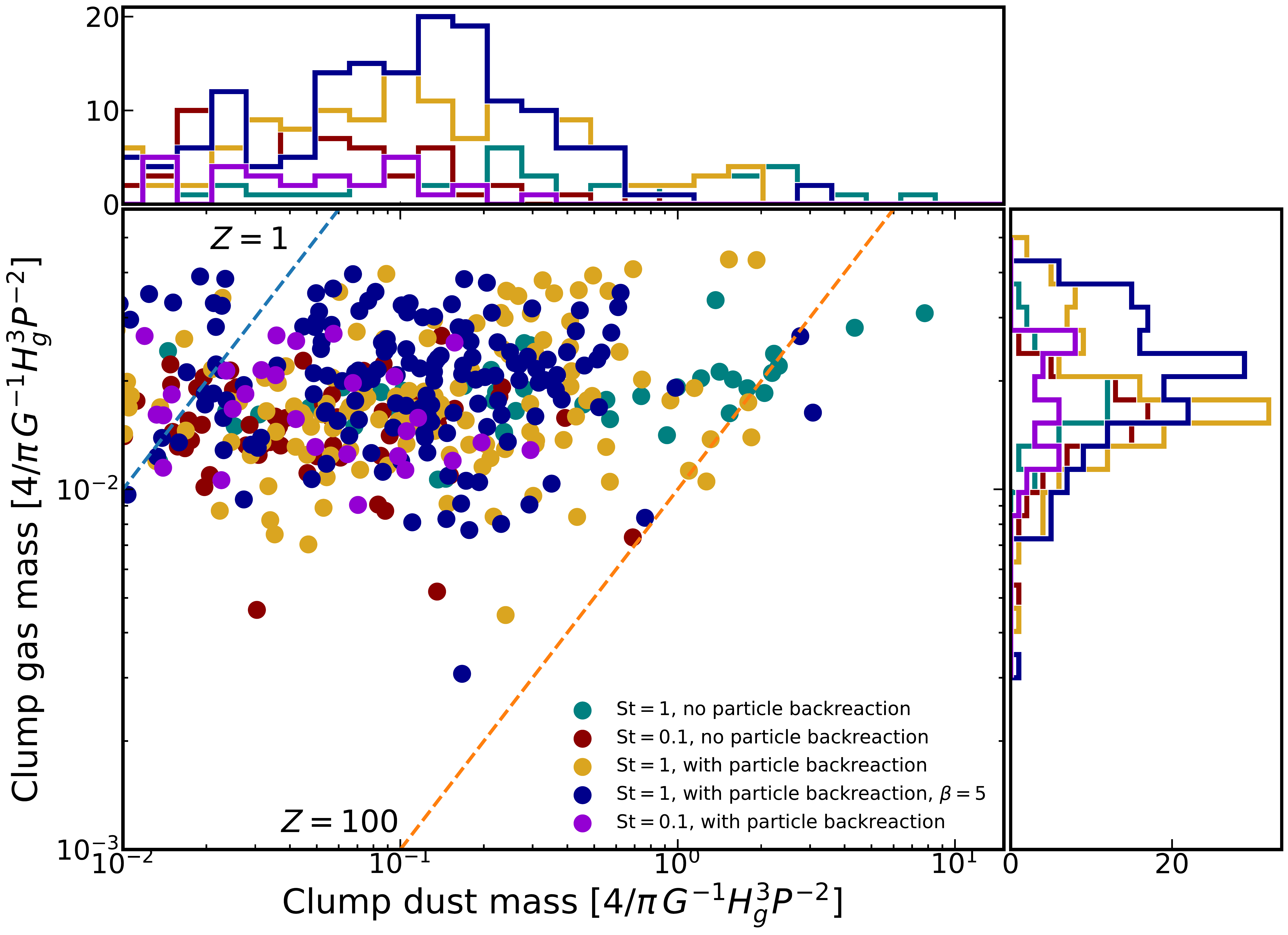}
\caption{Scatter plot of the total gas mass versus total dust mass within the Hill radius of each identified clump of particles. Dashed lines indicate constant dust-to-gas mass ratios. Plotted on top and to the right are the logarithmically-binned histograms of the data.}
\label{fig:clumpmetallicity}
\end{figure*}

At the smaller particle size, $\mathrm{St}=0.1$, particles couple more tightly to the gas and the flows at smaller scales. As shown in Figure \ref{fig:supersonicflows}, less efficient settling to the midplane by the $\mathrm{St}=0.1$ particles means gas flows near the surface can occasionally sweep up particles \citep{Riols2020,Baehr2021a}. While clumps form with smaller dust, the particle layer is not as well settled and the local dust loading is lower. We present the full 3D view of the gas and dust in Figure \ref{fig:3drender}, where we highlight the most massive clumps with gray circles.

The velocity distribution of the largest dust species ($\mathrm{St}=10$) is noticeably shifted towards higher velocities than the smaller particle sizes studied here. Even though there are only small differences between the gas velocity distributions in all simulations, particles of the largest dust species is more likely to be at higher velocities and even a significant amount in the supersonic regime. Larger dust species are less coupled to gas motions, and gravitational stirring from the dense gas filaments can excite the larger dust grains to much higher velocities \citep{Shi2016}.

We measure radial particle diffusion constants using the method described in Section~\ref{sec:gravitationalcollapse}. Our models have the same resolution in the radial and azimuthal directions as in the 2D simulations of \cite{Shi2016} and we derive a similar increasing trend in the diffusion constants up to $\mathrm{St}=1$. However, whereas 2D simulations showed that radial diffusion is highest when $\mathrm{St}=10$ and decreases with increasing particle size, we notice that radial diffusion is already beginning to decrease from $\mathrm{St}=1$ to $\mathrm{St}=10$ in our 3D models.

Similar to \citet{Baehr2021a}, vertical dust diffusion is inferred from the dust scale height once the simulation has reached an equilibrium and the dust scale height is quasi-steady in time. With the ratio of dust scale height and the equilibrium gas scale height due to the gas self-gravity $H = 0.75 H_g$, the vertical dust diffusion constant can be determined via (\citealt{Dubrulle1995}; see also \citealt{Yang2018})
\begin{equation}\label{eq:dustscaleheight}
\frac{H_{d}}{H} = \sqrt{\frac{\delta_{d,z}}{\delta_{d,z} + \mathrm{St}}},
\end{equation}
where $\delta_{d,z} \equiv D_{d,z} / c_s H_g$ (Section~\ref{sec:analysis}). The measured vertical particle diffusion constants are sometimes around two orders of magnitude lower that the radial particle diffusion, indicating a high degree of anisotropy in dust transport driven by gravitoturbulence.

Particle diffusion is measured with a cadence of $\Delta t=2\Omega^{-1}$ and reported in Table \ref{tab:sims}, along with the corresponding Schmidt number $\mathrm{Sc} \equiv (\alpha_R + \alpha_G)/ \delta_{d,z}$. The Schmidt number is a dimensionless number that measures how well coupled the dust diffusion is to the gas turbulence \citep{Dullemond2004}. The Schmidt numbers derived here are particularly high, as is often the case for self-gravitating turbulence \citep{Riols2020,Baehr2021a}. Schmidt numbers are notably higher in simulations where more gravitationally bound clumps are identified. This could in part be due to the large number of particles in clumps will skew the vertical particle distribution towards a narrower profile and lower measured diffusion constant. The simulation with the largest dust species has the lowest Schmidt number. However, low vertical diffusion could instead help cause the collapse of dust rather than be a symptom of particle collapse.

\subsection{Cloud Masses}
\label{subsec:massresults}

All simulations carried out here show a similar size distribution that is dominated by a larger number of smaller clumps with a small number of clumps that are larger by over an order of magnitude in mass. How each population arrives at this distribution differs both by size of the dust and whether backreaction is included.

We characterize the mass distribution of solids using an exponentially-tapered power law, modified to use minimum mass $M_{\mathrm{min}}$ as a fitting parameter rather than the characteristic mass $M_{\mathrm{pow}}$ as in \citet{Schafer2017}. We therefore define the number of clumps above a particular mass $N_{>}(M)$ as
\begin{equation} \label{eq:exptaperpowerlaw}
\frac{N_{>}(M)}{N_{\mathrm{tot}}} = \left( \frac{M}{M_{\mathrm{min}}} \right)^{-\alpha} \mathrm{exp}\left[ \left( \frac{M_{\mathrm{min}}}{M_{\mathrm{exp}}} \right)^{\beta} - \left( \frac{M}{M_{\mathrm{exp}}} \right)^{\beta} \right],
\end{equation}
and use it to fit the distributions at a time $t = 60\,\Omega^{-1}$ when gas stability parameter $Q$ is steady in time. This can be compared with other studies of planetesimal formation via gravitational collapse of particles \citep[e.g.][]{Johansen2015,Schafer2017,Abod2019,Li2019a}. We find that an exponentially tapered power law fits the data best, but not as well at higher masses, similar to the results of \citet{Johansen2015}, \citet{Schafer2017} and \citet{Abod2019}. \citet{Li2019a} finds the best fit distribution varies with simulation parameters, but a truncated broken power law, broken cumulative power law or three segment power law can all match the data.

We find a noticeable difference between clump formation with particle backreaction versus those without when particle size $\mathrm{St}=1$. Backreaction from particles causes more rapid concentration of the dust at the earliest onset of dust sedimentation and collapse. This creates a large number of smaller dust clumps that will only grow larger through hierarchical mergers of clumps. Without backreaction however, particles do not segregate into individual clumps as rapidly, instead forming long filament-like structures which will later collapse into fewer, but larger objects. This is shown in Figure \ref{fig:cumulativemassdistributionevolution}, where the simulations with dust backreaction have considerably more identified clumps at the lower mass end to the left, but fewer at the higher end of the distribution to the right.

\begin{deluxetable*}{cccccccccccc}
\tablecaption{Particle clump properties:\label{tab:clumps}}
\tablehead{
\colhead{model} & \colhead{$\overline{N}$} & \colhead{ \begin{tabular}[c]{@{}l@{}} $\overline{\langle M_{\mathrm{dust}}\rangle}$ \\  $[\hat{M}_{0}]$ \end{tabular}} & \colhead{ \begin{tabular}[c]{@{}l@{}} $M_{\mathrm{dust,max}}$ \\  $[\hat{M}_{0}]$ \end{tabular}} & \colhead{ \begin{tabular}[c]{@{}l@{}} $\overline{\langle M_{\mathrm{gas}}\rangle}$ \\  $[\hat{M}_{0}]$ \end{tabular}} & \colhead{$\overline{\langle Z \rangle}$} & \colhead{\begin{tabular}[c]{@{}l@{}} $\overline{\langle\sigma\rangle}$ \\ $[c_s]$ \end{tabular}} & \colhead{$\alpha$} & \colhead{$\beta$} & \colhead{\begin{tabular}[c]{@{}l@{}} $M_{\mathrm{min}}$ \\  $[\hat{M}_{0}]$ \end{tabular}}& \colhead{ \begin{tabular}[c]{@{}l@{}} $M_{\mathrm{exp}}$ \\  $[\hat{M}_{0}]$ \end{tabular}}}
\startdata
noBR\_S\_10 & $37$ & $0.11$ & $0.69$ & $0.12$ & $0.70$ & $3.2\times 10^{-2}$ & $-5.9\times 10^{-1}$ & $0.35$ & $4\times 10^{-3}$ & $9.4\times 10^{-4}$\\
noBR\_L\_10 & $37$ & $1.01$ & $7.8$ & $0.15$ & $4.8$ & $5.0\times 10^{-2}$ & $3.0\times 10^{-2}$ & $0.58$ & $5\times 10^{-3}$ & $0.74$ \\
BR\_S\_10 & $16$ & $0.11$ & $0.20$ & $0.14$ & $0.52$ & $1.5\times 10^{-1}$ & $6.0\times 10^{-1}$ & $2.5$ & $1\times 10^{-2}$ & $0.2$ \\
BR\_L\_10 & $115$ & $0.29$ & $1.9$ & $1.8$ & $10$ & $5.5\times 10^{-2}$ & $-1.1\times 10^{-1}$ & $0.51$ & $4\times 10^{-3}$ & $8.8\times 10^{-4}$ \\
BR\_XL\_10 & $0$ & -- & -- & -- & -- & -- & -- & -- & -- & -- \\
BR\_L\_5 & $164$ & $0.18$ & $3.1$ & $0.16$ & $1.3$ & $8.0\times 10^{-2}$ & $4.6\times 10^{-2}$ & $0.97$ & $5\times 10^{-3}$ & $0.21$ \\
noBR\_L\_10\_HR & $82$ & $0.31$ & $1.7$ & $0.11$ & $3.2$ & $1.0\times 10^{1}$ & $-4.5\times 10^{-1}$ & $0.36$ & $2\times 10^{-3}$ & $0.55$ \\
\enddata
\tablecomments{Summary of clump properties and particle properties within bound clumps in each simulation where $N$ is the number of identified bound objects, $M_{\mathrm{dust}}$ and $M_{\mathrm{gas}}$ are the dust and gas mass within the Hill sphere of each clump, $Z = M_{\mathrm{dust}}/M_{\mathrm{gas}}$, $n$ is the number of particles within a clump, $\sigma$ is the particle velocity dispersion, and $\alpha$, $\beta$, $M_{\mathrm{min}}$ and $M_{\mathrm{exp}}$ are best fit parameters in Equation \eqref{eq:exptaperpowerlaw}. Quantities in $\langle\cdot\rangle$ brackets are averaged over all clumps in the simulation. Quantities with a bar above are averaged over multiple snapshots in time from $t=60\,\Omega^{-1}$ to $t=80\,\Omega^{-1}$, except for the high resolution run which was calculated from $t=20\,\Omega^{-1}$ to $t=30\,\Omega^{-1}$. The simulation with $\mathrm{St}=10$ particles 'xlarge' did not form any bound clumps, but is included for completeness. Masses are in units of $\hat{M}_{0}=(4/\pi) G^{-1}H^{3}_{g}p^{-2}$ (Equation \eqref{eq:codemassunit}).} 
\end{deluxetable*}
The more efficiently drifting particles with $\mathrm{St}=1$ are especially efficient at clumping and form more clumps with higher masses. These clumps gradually merge over time, and the total number of clumps decreases while the number in the high mass end increases. This could suggest that planetesimal-planetesimal accretion is the more common way to grow than pebble accretion. However, the opposite appears to be the case when dust is dominated by smaller species. The size distribution of clumps with smaller particles ($\mathrm{St}=0.1$) evolves in a different way as illustrated in Figure \ref{fig:cumulativemassdistributionevolution}. Bound clouds of dust formed when $\mathrm{St}=0.1$ are initially scarcer and smaller than those at $\mathrm{St=1}$. The number of bound clumps does steadily rise across the entire mass range however. Since the number of objects is lower than in the case of $\mathrm{St}=1$ and more space between them, planetesimal-planetesimal mergers is less likely to be the cause of clump growth. Instead, the plentiful dust in and between the filamentary dust structures is likely accreted by the clumps.

Since the simulations presented here are very nearly 3D analogs of those in \citet{Gibbons2014a}, we compare our findings with theirs. Their most massive dust clouds have notably lower masses, nearly two orders of magnitude of lower than identified here. Additionally, while the number of clumps is not identified in their work, by visual inspection there appear to be fewer in the 2D simulations.

In Figure \ref{fig:clumpmetallicity}, we compare clumps the dust and gas masses at a single snapshot in all simulations. Dust masses vary by a few orders of magnitude, and tend to be broadly distributed when numbers are low. When large numbers of dust clumps form, as in the two simulations where dust particles have size $\mathrm{St}=1$ and include backreaction, clumps tend to have masses around $0.1 \hat{M}_0$. Gas masses consistently fall within a more narrow range, which makes sense when one considers that all were formed within the non-axisymmetric gas structures which do not vary strongly in peak gas density. The small handful of low gas masses are the result of clumps that have drifted out of the dense gas structure in which they formed. Overall, most dense particle regions have high dust-to-gas ratios between $Z=1$ and $Z=100$.

It is important to consider that resolution and initial conditions may have an effect on the results. Convergence is not a given, particularly in simulations that concern the gravitational collapse \citep{Meru2011b,Paardekooper2012}, potentially due to effects of dimensionality \citep{Baehr2017} or other numerical factors \citep{Deng2017,Klee2017,Klee2019}. Thus we performed a simulation with double the grid resolution of the noBR\_L\_10 simulation and compared the mass distribution. The number of particles was also increased in the high resolution run to maintain the same number of particles in a midplane layer assumed to be one cell thick, leading to a four fold increase in particles. The high concentrations of particles that lead to clumping could also be a result of the linear collapse phase of the gas. Therefore, we also conducted a simulation where particles were only added to the simulation after the gas had reached a turbulent steady state.
\begin{figure}[t]
\centering
\includegraphics[width=0.45\textwidth]{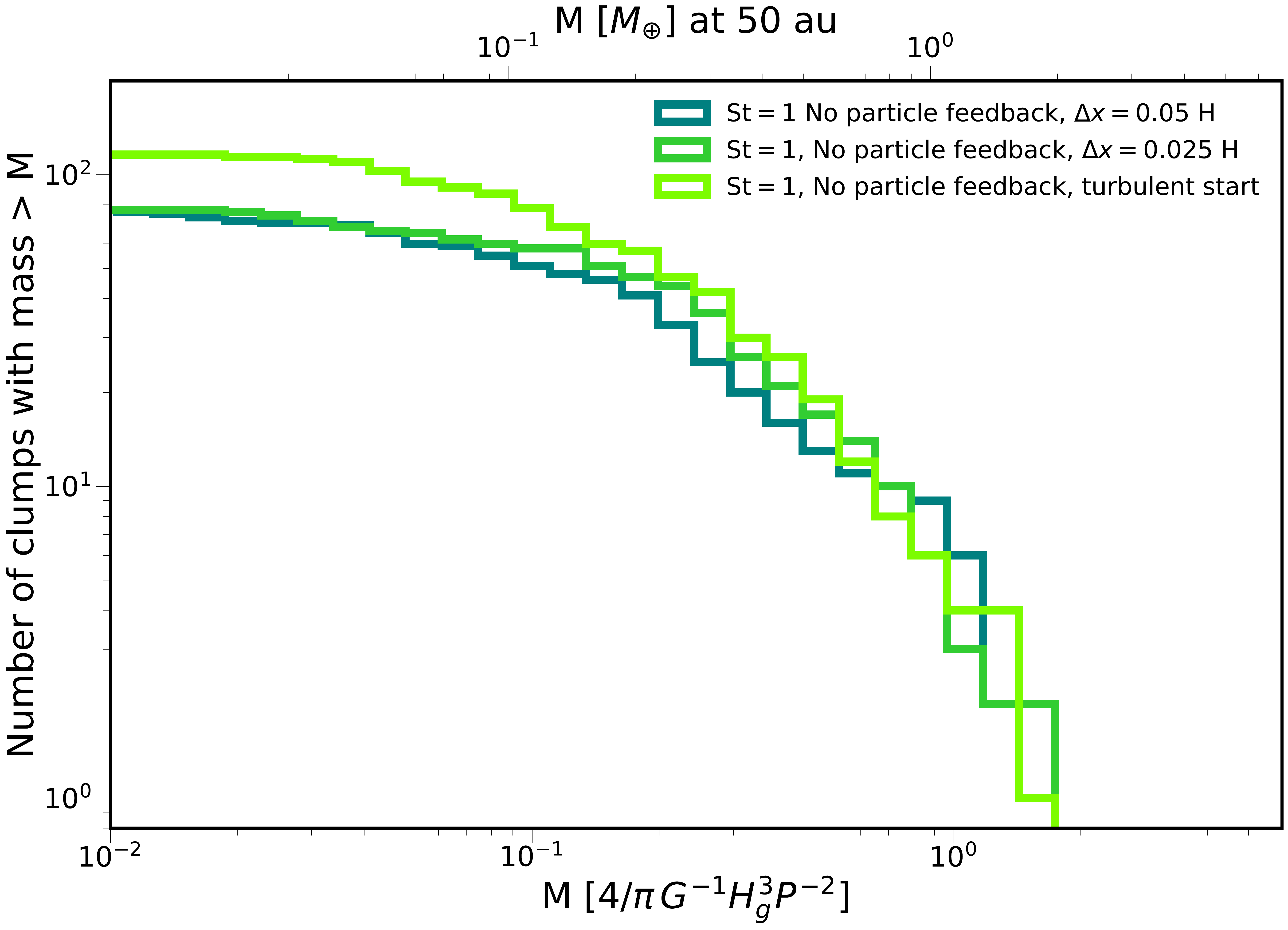}
\caption{Comparison of the mass distribution of clumps in simulations with the same parameters but with different resolution and initial conditions}. All mass distributions are from the same point in time after particle are introduced for $t = 26\,\Omega^{-1}$.
\label{fig:clumpconvergence}
\end{figure}

Figure \ref{fig:clumpconvergence} shows the distributions of all three simulations at the same time $t = 26\,\Omega^{-1}$. In each case, the upper clump mass limit is the same and the number of identified bound clumps is similar, with more smaller clumps forming in the simulation where dust particles were added later. Clump and turbulent diagnostics of the high resolution simulation are reported in Tables \ref{tab:sims} and \ref{tab:clumps} and are within reasonable values compared to the standard resolution. Simulations of planetesimal formation via the streaming instability also appear similarly converged with resolution, such that the differential size distribution is unaffected by grid resolution \citep{Johansen2015}.

It should be noted that this suggests convergence for the simulations where the dust scale height is resolved ($\mathrm{St} = 0.1$ and $\mathrm{St} = 1$). For $\mathrm{St} = 10$, increasing the resolution should resolve the critical wavelength of the dust and change the results at this size (see Section~\ref{subsec:particlecollaspe}).

\section{Discussion}
\label{sec:discussion}

\subsection{Particle Collapse Criterion}
\label{subsec:particlecollaspe}

\begin{figure}[t]
\centering
\includegraphics[width=0.45\textwidth]{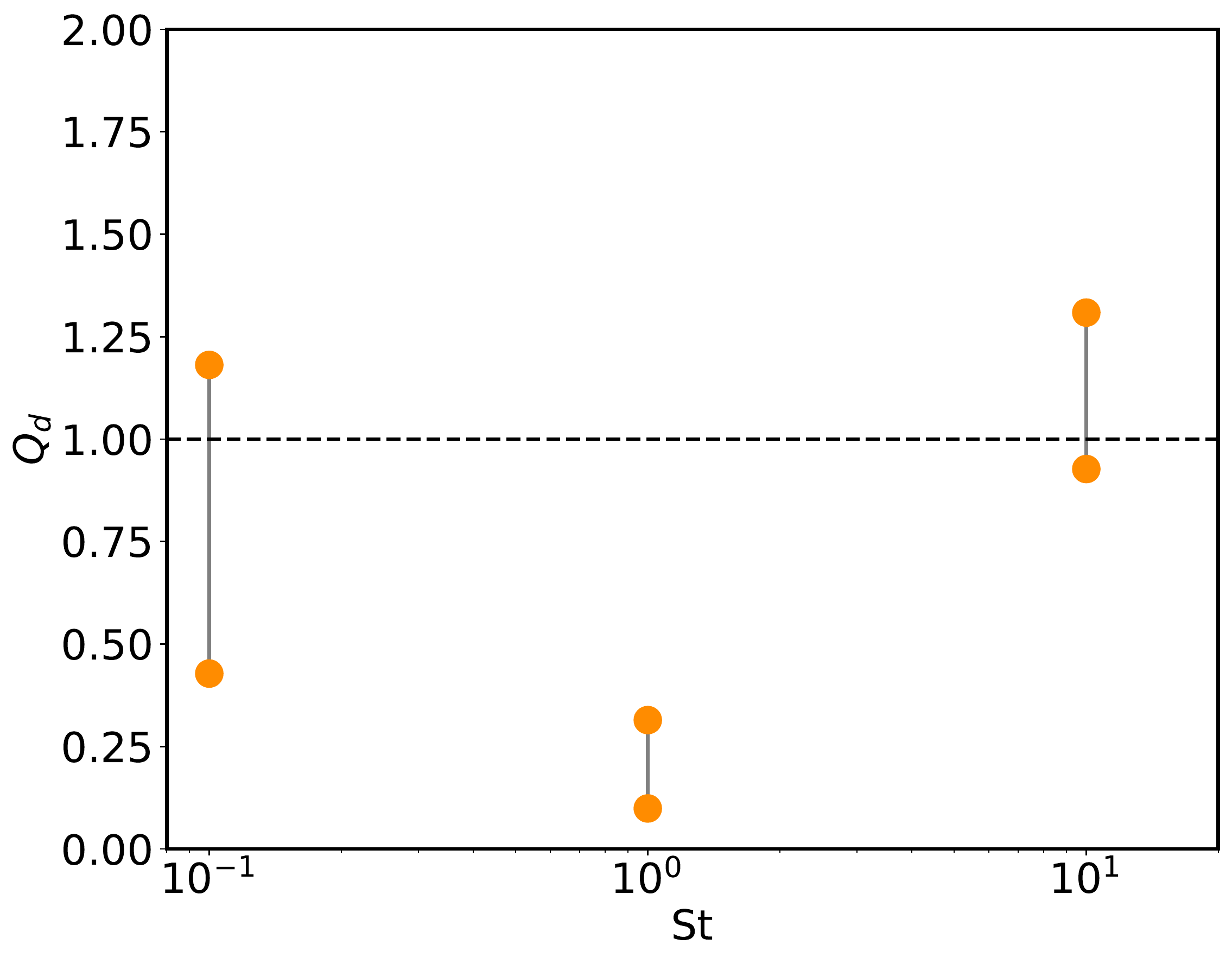}
\caption{Particle stability parameter $Q_{d}$ for three particle sizes which include the effect of particle backreaction. The diffusion constant and range of maximum concentration $\epsilon Z$ in Equation \eqref{eq:gerbigparameter} is determined from the $\Sigma_{d,\mathrm{max}}/\langle\Sigma_{d}\rangle$ across a few snapshots after concentration in the spiral structures has occurred but before particle self-gravity dominates the high end of particle densities ($t=12$--$16\,\Omega^{-1}$).}
\label{fig:particleq}
\end{figure}

The particle collapse criterion of equation~\eqref{eq:gerbigparameter} from Section \ref{sec:gravitationalcollapse} can be tested for the three particle species $\mathrm{St} = 0.1,1,10$. Because particle self-gravity significantly affects the particle density enhancement ($\epsilon Z$) once gravitational collapse is triggered, the criterion has to be measured after concentration into filaments has started, but before particles have collapsed. For this purpose, we consider the time span $t \in [12\Omega^{-1}, 16\Omega^{-1}]$ to measure $\epsilon Z$ as well as the radial diffusion constants $\delta_{x}$ to compute $Q_{\mathrm{d}}$ for the simulations that include particle backreaction.
The resulting range of $Q_{\mathrm{d}}$ in this duration are plotted in Figure \ref{fig:particleq}.

The two smaller sizes of dust species reach values of $Q_{\mathrm{d}}$ well below the stability threshold. Considering the simulations have an equilibrium gas stability parameter of $Q \sim 1.2 - 1.4$, suggesting that dust particles can become gravitationally unstable and collapse even when the gas disk is marginally stable. If disk becomes more unstable with time through the accretion of envelope material \citep{Kuffmeier2018,Kuznetsova2022}, structure in massive/gravitoturbulent disks may be determined not only by the instability of the gas, but the dense solid concentrations that form first. On the contrary, due to higher radial dust diffusion, the simulation with the largest dust particles straddles the stability threshold and the particles in the simulation do not collapse into dense clumps.

When the simulation reaches a quasi-steady state after gravitational collapse of dust, the diffusion decreases. The dust diffusion can decrease to the extent that the critical length scale of the dust is no longer resolved and gravitational collapse at this size cannot occur. The critically unstable length \citep{Gerbig2020,Klahr2020}
\begin{equation}
r_{\mathrm{crit}} = \frac{1}{3} \sqrt{\frac{\delta_{x}}{\mathrm{St}}} H_{g}
\end{equation}
for $\mathrm{St}=10$ is $r_{\mathrm{crit},10} = 0.01 H_{g}$ in the turbulent steady state, which is smaller than the grid spacing, indicating that there is not enough resolution to capture collapse at this size. Even at $\mathrm{St}=1$, $r_{\mathrm{crit},1} = 0.05 H_{g}$ and hence this dust species is not well resolved, but the dense clumps have already formed by this point from the initial collapse. Thus to study the collapse of dust sizes $\mathrm{St}=1$ and larger, grid spacing will have to be nearly an order of magnitude smaller. At smaller dust sizes, on the other hand, the critical length is resolved by a few grid cells, albeit not all unstable lengths are resolved.

Additional diffusive processes, such as the ones driven by the streaming instability and Kelvin-Helmholtz instability, can prevent collapse at scales smaller than a scale height, and the particle collapse criterion can be applied in the same way \citep{Gerbig2020}. While scales smaller than $\sim 0.1 H_g$ are not resolved in our simulations to capture SI or KHI, the high clump metallicities would favor low values of $Q_{\mathrm{d}}$ and the formation of planetesimals or planetary embryos. Although SI and KHI are unlikely to compete with GI to determine the stability of dust clumps, the interaction of three together should be explored in future works.

The stability of dust sizes outside the range of this study, in particular the smaller sizes, should be explored in future work. Self-gravitating disks are likely to be young, with less grain growth and coagulation to larger sizes, so understanding the lower particle size range susceptible to gravitational collapse will be important. Due to timestep limitations at dust sizes smaller than $\mathrm{St}=0.1$, we do not include any simulations at these small sizes.

\subsection{Clump Survival}
\label{subsec:clumpsurvival}

As shown, particles not only settle to the midplane, but they drift to the dense gas structures generated in a $Q \approx 1$ disk and collapse of dust into gravitationally bound clouds follows shortly thereafter. The power spectrum of the turbulent eddies of gravitoturbulent disks peaks at scales around the gas scale height \citep{Cossins2009,Booth2019} and decreases down to the sizes of the clumps. Interactions with the high velocity gas motions can potentially push some dust clumps into supersonic regime. This is supported by the high collective motions measured for a few clumps in our simulations, seen in Figures \ref{fig:gasdustvelocities} and \ref{fig:gasdustvelocities2}. While gas velocities can often reach the supersonic regime close to the midplane but rarely in the midplane, the clumps are pushed collectively but particle dispersions rarely exceed the escape velocity.

\begin{figure}[t]
\centering
\includegraphics[width=0.45\textwidth]{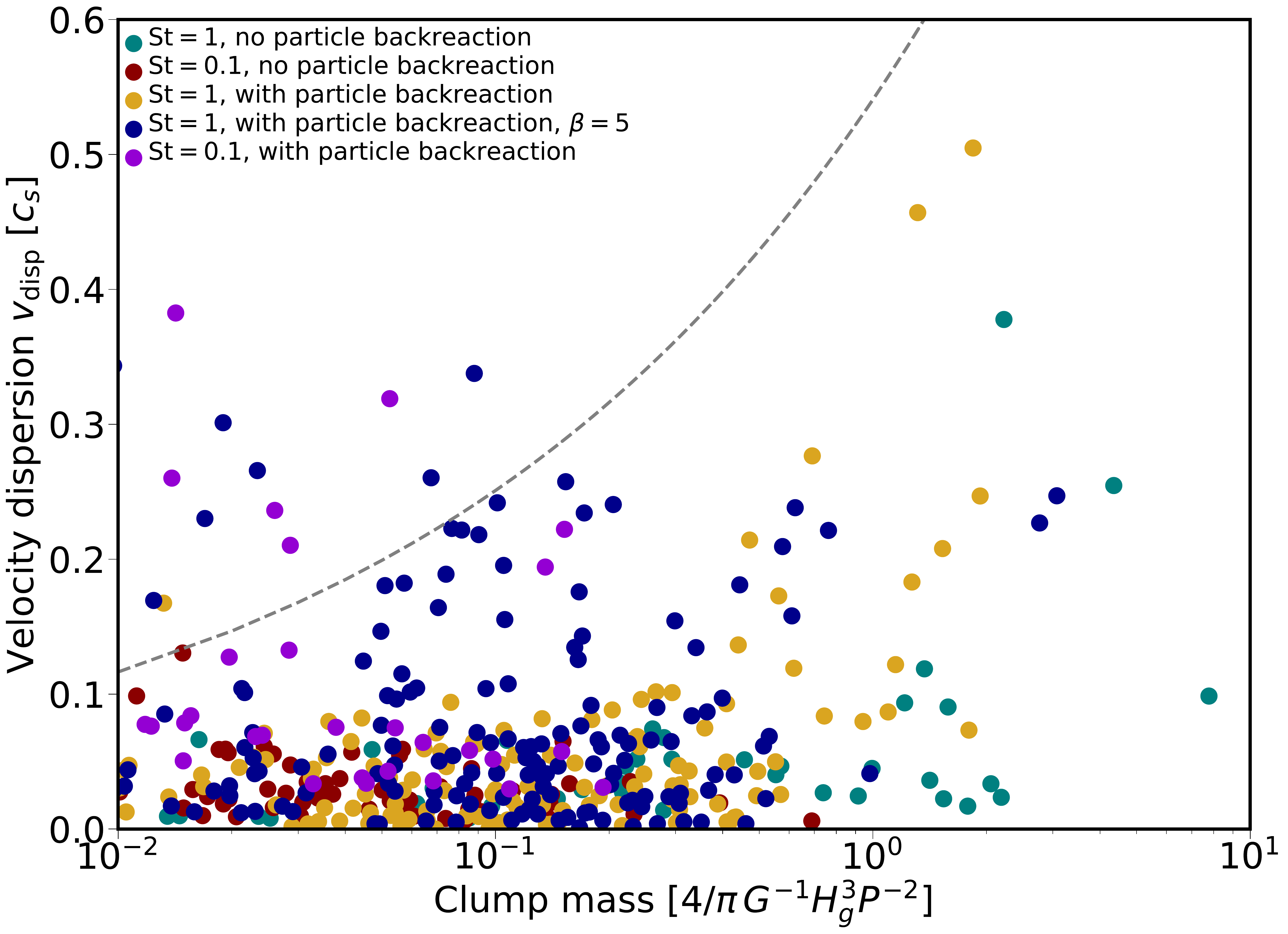}
\caption{Velocity dispersion of particles in a clump versus the dust mass of the clump. The gray dashed line indicates the threshold where the particle velocity dispersion within the clump is the local escape velocity. A particle clump above this line will likely be disrupted if it does not accrete more particles or the particles in the clump become less excited.}
\label{fig:clumpvelocitydispersion}
\end{figure}

In Figure \ref{fig:clumpvelocitydispersion}, we measure the particle velocity dispersion in each clump. The velocity dispersion with any given clump is generally less than 10\% the sound speed, but a small number of clumps at all masses have dispersions approaching $0.5 c_{s}$. The escape velocity $v_{\mathrm{esc}}$ of a particle from a massive cloud can be found in terms of the gas sound speed through the definition of the Bondi radius
\begin{equation}
\left( \frac{v_{\mathrm{esc}}}{c_s} \right)^2 = \frac{2GM_{\mathrm{c}}}{c_s^2 r} = \frac{2R_B}{r},
\end{equation}
where $M_{\mathrm{c}}$ is the mass of the clump and $r$ is the distance away from the center of the clump. For particles within the Hill radius $r \sim R_{H}$, we arrive at an expression for the minimum escape velocity in terms of the clump mass as a fraction of the thermal mass \citep[see][]{Fung2019a} and our unit mass $M_{\mathrm{th}} = M_{*}(H_{g}/R)^{3} = \pi^3\hat{M}_0$
\begin{equation}
\frac{v_{\mathrm{esc}}}{c_s} = 2^{1/2}3^{1/6} \left( \frac{M_{\mathrm{c}}}{ M_{\mathrm{th}}} \right)^{1/3} = 0.54 \left( \frac{M_{\mathrm{c}}}{\hat{M}_{0}} \right)^{1/3}.
\end{equation}
As shown in Figure \ref{fig:clumpvelocitydispersion}, some low mass clumps are above this threshold and contain particles that are moving fast enough to fall apart, while the majority of clumps either have sufficiently low particle velocity dispersions or high enough masses to remain bound. This indicates that these clumps are not just transient consequences of turbulent concentration, but persistent and stable clouds of particles.

\subsection{Planetesimal and Embryo Formation}
\label{subsec:embryoformation}

The formation of early planetary embryos and potentially planetary cores could have important implications for the formation of gas giant planets. Planets potentially carve out the gas and/or dust gaps on wide orbits observed in ALMA images of young systems. Systems younger than one million years \citep[i.e.][]{Alves2020,Segura-Cox2020} already show evidence of structures that could be the result of planet formation. This would work together with pebble accretion models, which require a solid seed mass of on the order of a fraction of an Earth mass \citep{Bitsch2015,Bitsch2019,Tanaka2020}. The accretion of additional pebbles and gas could take a few hundred to a million years \citep{Andama2022}.

The streaming instability is an efficient mechanism to form planetesimals in the inner few tens of au in protoplanetary disks, but less efficient at more distant orbital separations. Thus, gravitoturbulent disks provide a means to concentrate dust early such that the gravitational collapse could produce solid bodies up to the size of a few Earth masses at several tens of au early in the disk lifetime.

Gravitotrubulent disks are not the only way to form early planetesimals. Besides the aforementioned streaming instability, secular gravitational instabilities \citep{Tominaga2020} could fill the same role in a young disk while dust trapping in vortices via hydrodynamic instabilities could form the seeds of early planet formation \citep{Lyra2008,Raettig2015,Lyra2019}.

\subsection{Limitations}
\label{subsec:limitations}

Without a radial pressure gradient, there is no radial drift included in these simulations, which would add significant radial velocities to all particles, but particularly the $\mathrm{St}=1$ particles. A radial pressure gradient could also induce the streaming instability which would add further dust diffusion. What effects radial drift ultimately has on the ability of dust particles to concentrate should be further unexplored in 3D stratified gravitoturbulent simulations. 

While fully able to resolve the required length scales for gravitational collapse of the gas, ideally these simulations would be able to resolve the midplane dust layer with more than $\sim 10$ grid cells. Even more helpful would be to model GI and particle-gas instabilities in the same simulation such that one can continue the gravitational collapse of the dust down to smaller scales. Shearing boxes with adaptively refined grids will be the most useful tool to explore this scenario.

We considered grains all to be of the same size, which are also sizes most likely to drift into the dense gas structures formed by GI. This is in part for simplicity, but also because the interaction of multiple concurrent species through backreaction effects is not yet fully understood \citep{Zhu2021,Yang2021}. In reality, dust should initially have a size distribution weighted heavily towards smaller grains similar to the interstellar medium, where not much growth has occurred. The dust size distribution in the ISM scales with size as a power law to the $-3.5$ \citep{Mathis1977}. This would likely disfavor the mass distributions that rely on the more massive dust particles, but dust growth from $\mu$m to mm sizes can be efficient once the disk has formed \citep{Birnstiel2016}. Future work will have to determine the mass of particle clouds that can be formed when using a realistic dust size distribution. On the other hand, gravitationally unstable disks are most likely to occur directly after star formation \citep{Kuffmeier2018,Xu2021}, which may limit the amount of time for the collisional growth of dust up to the sizes included in this work. Whether collisional growth in these disks is efficient enough to create enough large dust grains to reach the point where the particle collapse criterion applies is a topic of continued research \citep{Sengupta2019,Elbakyan2020}.

\section{Conclusion}
\label{sec:conclusion}

Gravitationally unstable disks are dominated by their gas content but may concentrate significant quantities of solid material via gravitoturbulence which triggers the self-gravitational collapse of concentrated dust. We use 3D hydrodynamical simulations to compare how dust particles of different sizes concentrate, collapse, and form bound objects. In these simulations, we consider both simple dust drag without backreaction from the dust and a self-consistent drag backreaction from the dust onto the gas. This helps to better understand how dust and gas velocities are affected by one another and how the mass distribution of bound solid dust clouds changes in each case. We summarize the results of the paper below:

\begin{enumerate}
\item Including self-gravitating dust in 3D simulations of disk gravitoturbulence produces bound clumps of dust up to several $M_{\oplus}$ in mass at $\sim$50 au. Larger clumps are most efficiently formed when dust is of size $\mathrm{St}=1$, while for smaller sizes $\mathrm{St}=0.1$, clumps form but with a maximum mass roughly an order of magnitude lower.
\item The formation of clumps at the small and intermediate sizes is consistent with a particle collapse criterion based on the disruption of internal dust diffusion. No bound clumps form when particles are size $\mathrm{St}=10$, which is likely due to insufficient resolution of the critically unstable length at this size.
\item The effect of dust backreaction onto the gas can increase the perturbed velocities of both dust and gas into the supersonic regime. The clumps as a whole move at these high velocities, but the particle velocity dispersion within the clumps remains low and the clumps are stable over extended periods of time.
\item The formation of solid objects as early as the period when a circumstellar disk is self-gravitating could provide the seeds for planet formation. Growth of these embryos via pebble accretion models could help to explain the existence of early ring/gap structure in protoplanetary disks. 
\end{enumerate}

\acknowledgments
HB thanks Hubert Klahr and Konstantin Gerbig for valuable discussions and comments. Simulations were conducted on the Pleiades supercomputer hosted by the NASA Advanced Supercomputing (NAS) division as a part of the High-End Computing (HEC) program. This research was supported by NASA Theoretical and Computational Astrophysics Networks (TCAN) award 80NSSC19K0639 and discussions with associated collaborators.
CCY and ZZ acknowledge the support by NASA via the Emerging Worlds program (grant number 80NSSC20K0347) and the Astrophysics Theory Program (grant number 80NSSC21K0141). 
CCY is also grateful for the support by NASA TCAN program (grant number 80NSSC21K0497). ZZ acknowledges support from the National Science Foundation under CAREER Grant Number AST-1753168.

\software{Matplotlib \citep{Hunter2007}, SciPy \& NumPy \citep{Virtanen2020,vanderWalt2011}, IPython \citep{Perez2007}}

\bibliography{library.bib}

\end{document}